\documentclass[12pt,preprint]{aastex}
\newcommand{\etal}{{et al.}}
\newcommand{\eg}{{\it e.g.,}}
\newcommand{\ie}{{\it i.e.,}}

\begin{document}

\title{Quasi-Stellar Objects, Ultraluminous IR Galaxies, and Mergers\footnotemark[1]}

\footnotetext[1]{Based on observations with the NASA/ESA Hubble Space Telescope,
obtained from the data archive at the Space Telescope Science Institute, 
which is operated by the Association of Universities for Research in Astronomy,
Inc. under NASA contract No. NAS5-26555.}

\author {Gabriela Canalizo\altaffilmark{2}}
\affil{Institute for Astronomy, University of Hawaii, and the Institute of Geophysics and Planetary Sciences, Lawrence Livermore 
National Laboratory, 7000 East Avenue, L413, Livermore, CA 94550}

\author {Alan Stockton\altaffilmark{2}}
\affil{Institute for Astronomy, University of Hawaii, 2680 Woodlawn
 Drive, Honolulu, HI 96822}

\altaffiltext{2}{Visiting Astronomer, W.M. Keck Observatory, jointly operated
by the California Institute of Technology and the University of California.}

\begin{abstract}
We test the hypothesis that QSOs are formed via
strong tidal interactions or mergers, initially going through an ultraluminous 
infrared phase.   Our approach is to look for traces to this phase 
in the host galaxies of QSOs.   We select a 
sample of low-redshift bona fide QSOs that may be in a
transitionary stage between ULIGs and QSOs.   These objects, which we shall 
call transition QSOs,  have an intermediate
position in the far infrared color-color diagram between the regions
occupied by the two classes of objects.
We carry out a systematic spectroscopic and imaging study of these objects
in order to determine their interaction and star-forming histories.
By modeling the spectra, we obtain ages for the recent
starburst events in the host galaxies and interacting companions.   
We have discussed in detail the first 5 objects in the sample in
previous publications; here we present results for the remaining
4 objects, and discuss the sample as a whole.
We find that all 9 transition QSOs are undergoing tidal interactions
and that 8 are major mergers. Every object also shows strong recent
star-forming activity, and in at least eight cases this 
activity is directly related to the tidal interaction.  The ages we
derive for the starburst populations range from currently active star 
formation in some objects, to post-starburst ages $\lesssim300$ Myr 
in others.  There is also a clear connection 
between interactions, starbursts, and QSO activity.
Seven of the QSOs in the sample are also 
ULIGs;  statistical considerations show that the two phenomena 
are necessarily physically related in these objects.  Our results
imply one of two scenarios: (1) at least some ULIGs evolve to
become classical QSOs, and the transition stage lasts $\lesssim300$ Myr,
or (2) at least some QSOs are born under the same conditions as ULIGs,
and their lifetime as QSOs lasts $\lesssim300$ Myr.
We discuss other properties and trends found in the sample, and propose
a model that accounts for all of them, as well as the youth of these
systems.
\end{abstract}

\keywords{galaxies: interactions --- galaxies: infrared --- 
galaxies: starburst --- galaxies: evolution --- quasars: individual 
(IRAS\,00275$-$2859, \\ IRAS\,04505$-$2958, PG\,1543+489, and I\,Zw\,1)}

\section{Introduction}

It is now widely accepted that the energy source powering QSOs
is gravitational in nature and that it involves some form of accretion 
of matter onto a compact massive object in the center of a galaxy, as first 
suggested by \citet{lyn69}.
But it has long been clear that such an object will fairly quickly
swallow up most of the material that comes within its vicinity
and normal relaxation processes repopulate such orbits far too slowly
to sustain the power output of a QSO.   We are left with the problem 
first clearly posed by \cite{gun79}: how does one feed 
the monster?  As if anticipating the problem, \citet{too72}
suggested the importance of interactions and mergers 
as fueling mechanisms in the section ``Stoking the Furnace?'' of their 
classic paper ``Galactic Bridges and Tails'', where they muse: ``Would not 
the violent mechanical agitation of a close tidal encounter---let alone an 
actual merger---already tend to bring {\it deep} into a galaxy a fairly 
{\it sudden} supply of fresh fuel in the form of interstellar material, 
either from its own outlying disk or by accretion from its partner?''  
Toomre \& Toomre's suggestion was specifically in reference to galaxies
showing vigorous nuclear star formation.   
However, it was only a small step to apply the same scenario to QSO activity,
a connection explicitly made by \citet{sto82} in a study of compact
companions to QSOs.

Many attempts to connect most or all QSO activity with interactions and 
mergers followed, and circumstantial evidence supporting
this connection was rapidly accumulated.
A variety of morphological 
distortions were seen in the 
host galaxies, ranging from simple asymmetries to clear bridge-like
or tail-like structures \citep[\eg][]{geh84,hut84,hut88a,sto87}.
There seemed to be a statistical excess of close companion galaxies
\citep[\eg][]{hec84}.
Highly structured distributions of  extended ionized gas were found around QSOs
\citep[\eg][]{bor85,sto87}.
\citet{sto90} reviews and evaluates these and other examples
and points out that what keeps these observations from having more weight
as evidence for a connection between interactions and the triggering of
QSO activity is the lack of a suitable control sample.

It was in this context that the {\it Infrared Astronomical 
Satellite} ({\it IRAS}) made the remarkable discovery of 
a new class of galaxies, the luminous infrared galaxies, which emit
the bulk of their energy at infrared wavelengths
\citep{hou84,soi84a,soi84b}.
Many of these galaxies showed unambiguous signs of tidal interaction
\citep{all85,san86,arm87},
and the interaction rate was observed to increase with luminosity.
In particular, virtually all of the ultraluminous infrared 
galaxies (\ie\ those having luminosities $> 10^{12}$ L$_{\sun}$; henceforth
ULIGs), were found to be involved in mergers (see \citealt{san96} for a
review).

Up until the time of {\it IRAS}, QSOs were the only objects known to have 
bolometric luminosities greater than $10^{12}$ L$_{\sun}$,   
so it was natural to speculate about 
about a relation between QSOs and the newly discovered ULIGs.  
\citet{san88} noticed that ULIGs and QSOs had not only similar bolometric
luminosities, but also similar space densities.
In other words, ULIGs were numerous enough to be the parent population
for QSOs, and their infrared luminosities could simply be due to
QSO flux being re-radiated at longer wavelengths.
This led \citet{san88} to suggest that ULIGs may play a dominant role in the
formation of all QSOs.  The picture put forward
was that ULIGs were the result of strong interactions or mergers
which funneled gaseous material into the central regions of galaxies,
thus fueling intense star formation and the QSO activity.  
ULIGs were then dust-enshrouded QSOs which, after blowing away
the dust, became classical QSOs.

This hypothesis has sparked much debate regarding the main
source that powers ULIGs: is it powerful starbursts \citep{jos99} or
AGNs \citep{san99}? 
\citeauthor{smi98} \citetext{1998; see also \citealt{lon93,lon95}}
have looked for the presence or absence
of a true high brightness temperature AGN-like radio core in ULIGs; 
\citet{gen98} and \citet{lut98} use the strength of the 7.7 $\mu$m
PAH line to continuum ratio to determine the fraction of the 
infrared/submillimeter luminosity peak in spectral energy distributions
(SEDs) that is due to a starburst and an AGN respectively, and \cite{lut99}
have compared the dominant contributor from this analysis with the
classification from optical/IR emission lines, finding general agreement; 
\citet{nak99} have used {\it ASCA} hard x-ray observations to estimate
the AGN contribution to the total luminosity; \citet{hin95a,hin99a},
\citet{goo96} and \citet{tra99}, among others, 
have looked for hidden QSO broad emission lines in scattered polarized light 
in ULIGs.  These are only a few examples of the myriad of tests at virtually 
every observable wavelength that have been applied to ULIGs with the goal
of determining the nature of the main energy source.   
Although some uncertainty remains, it appears that $\sim30$\% of ULIGs
are dominated by AGNs \citep{lut99} and that this fraction rises at the
highest luminosities \citep{vei99a}.

The limited attempts to address the question of evolution from the other
side of the fence (in other words, to look at QSO host galaxies 
specifically in search for traces of a ULIG phase) pale in
comparison.  Progress in this area has been damped by the difficulty 
in studying QSO host galaxies, as observations are invariably hampered by 
the enormous brightness of the QSO nuclei.   It is not too surprising that
even {\it Hubble Space Telescope} ({\it HST}) observations of host
galaxies caused some confusion at first.  \citet{bah95} originally presented a
scandalous set of ``naked'' QSOs, claiming that no galaxy as
bright as $L^{*}$ could be detected in 5 out of a sample of 8 low-redshift
QSOs.   Later, McLeod \& Rieke (1995) showed that, with suitable smoothing, 
$L^{*}$ galaxies were indeed detectable on the same {\it HST} images,
and concluded that most QSOs are in normal, early-type galaxies.
Efforts in the last decade have concentrated on characterizing
QSO host galaxies and, in particular, searching for any possible 
differences between the hosts of radio-loud and radio-quiet objects
\citep[\eg][]{mcl99,bah97,boy98},
as well as continuing to explore the connection to interactions and
mergers as means to trigger the nuclear activity 
\citep[\eg][]{hut94,mcl94,dis95,hut92,bah97,boy96}.
Nevertheless, all the 
evidence connecting QSOs to mergers remains {\it circumstantial}.

Systematic studies to attempt to trace QSOs to ULIGs have been few and 
limited to imaging studies of host galaxy morphologies.   \citet{sur98a}
observed a complete sample of PG QSOs with infrared excesses similar to 
those of warm ULIGs.   He finds that 22\% are unmistakably the result of 
galaxy mergers, and that the $H$-band luminosities of the hosts of all
the objects in the sample are consistent with those found in ULIGs.
\citet{cle00} compares the far infrared properties of QSOs in disturbed
and undisturbed hosts in the PG survey, and finds that the mean 60~$\mu$m
luminosity of the QSOs with disturbed hosts is several times greater than 
that of QSOs with undisturbed hosts.  This result would be in agreement with
a scenario where the disturbed-host QSOs are at an earlier stage in their
evolution from ULIGs to classical QSOs.    

These and other efforts have 
concentrated on the connection between QSOs and ULIGs via morphologies
indicative of mergers.   An area that has been virtually ignored, except
for the occasional qualitative comment, is the detailed study of the potential
connection between
both families via {\it starbursts}.   While the starburst--AGN connection
itself has been the subject of much recent research, no systematic
effort has been made to attempt to detect such starbursts in the host galaxies
of QSOs.   In particular, relatively few spectroscopic observations of QSO host
galaxies have been published 
\citep{bor82a,bor84,bor82b,bal83,mac84,bor85,hic87,hut90}.
The majority of these have been limited to
reporting the presence of absorption lines in the spectra of host galaxies, 
and most studies have been of isolated cases.  Again, this is not
surprising considering the difficulties in observing a low surface
brightness object only a few arcseconds (in the best of cases) away
from its $\sim1000$ times brighter nucleus.
One group \citep{hug00,nol00} is currently carrying out a systematic 
spectroscopic study of radio galaxies and both radio-loud and 
radio-quiet QSOs, but their
emphasis is in characterizing the old populations (while treating any possible
recent starburst components as a nuisance) and describing differences
among the hosts of the three classes of objects.
Thus, it is in this mostly unexplored territory of spectroscopy of QSO
host galaxies that we have focussed our efforts.

Our approach to the question of evolution is as follows:  in the 
evolutionary scenario, objects are expected to transition at some point 
from an ultraluminous infrared phase to
a ``normal'' QSO phase; we shall refer to objects in this latter phase
as ``classical QSOs.''   
Our intent is to define a group of ``transition'' objects, \ie\ objects
that share characteristics with both QSOs and ULIGs, and,
through both imaging and spectroscopic observations, answer the  
questions: {\it ``Where did they come from?''}, and {\it ``Where will 
they go?''} 
If the answers to these 
questions were unambiguously determined to be,  respectively, 
``from the ULIG population,'' and ``to the classical QSO population'', the 
evolutionary scenario would be confirmed for at least one class of
QSOs, and statistical considerations could then tell us whether the
process is relevant to the majority of QSOs.
In the following section we address the problem of
how to come up with a sample of transition objects,
describe our selection criteria, and present our sample.  
Results for the first 5 objects in the sample have been published in
\citet{can97}, \citet{can00a}, \citet{can00b}, and \citet{sto98} 
(hereafter CS97, CS00a, CS00b, and SCC98, respectively).
In \S\ref{chap5} we analyze and present results for the last 4 objects in 
the sample:
IRAS\,00275$-$2859, IRAS\,04505$-$2958, PG\,1543+489, and I\,Zw\,1.
In \S\ref{agesum} we summarize the observed and derived properties of all 
the objects in the sample, in \S\ref{properties} we discuss these properties 
individually, and in \S\ref{together} we present a model that accounts for 
all of them in the context of our main results.   
We discuss other possible tests of the evolutionary paradigm in 
\S\ref{further} and present an overview and summary in \S\ref{summary}.

\section{Sample Selection \label{samplesec}}

It is possible to select a sample of objects that
may be in some transitionary phase by choosing a set of bona fide QSOs that
share a given characteristic with ULIGs.   We could, for example, choose 
those QSOs that have recent signs of interaction.   However, the selection
of such a sample would be very subjective, and it would be virtually 
impossible to collect a complete sample and/or to come up with a suitable
control sample.  In an effort to avoid some of these problems, we have 
chosen to use far infrared colors to form a nearly complete sample.

\begin{figure}[p]
\epsscale{1.0}
\plotone{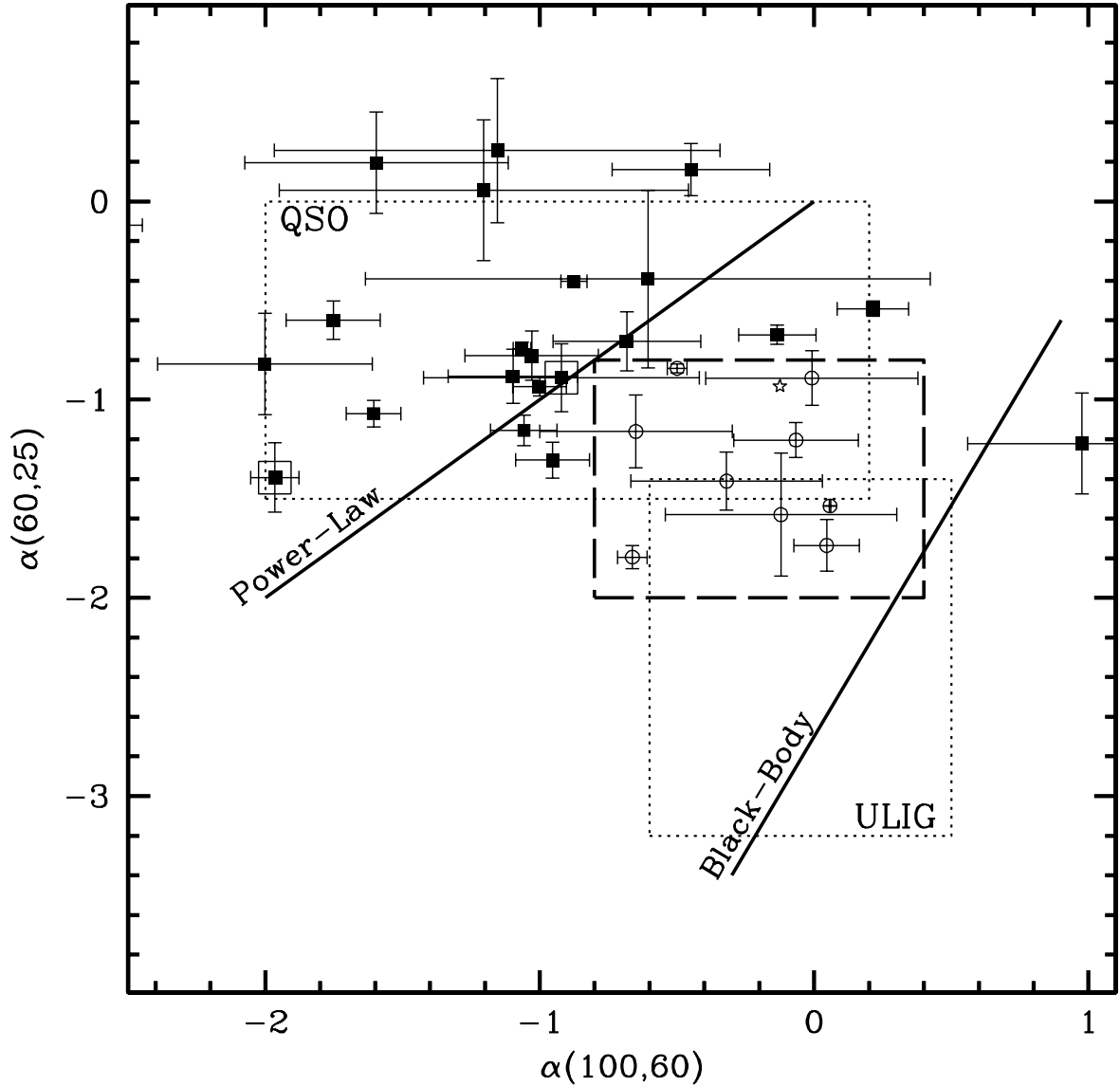}
\caption {Far infrared color--color diagram (adapted from L\'{\i}pari 1994).
Every object from the three samples
described in the text that satisfies selection criteria 1 through 4 is
plotted in this diagram.   The color limits for our sample of transition
QSOs are indicated
by the long dashed lines.  Objects in our sample are plotted as open circles
while all others are plotted as filled squares.  
The two filled squares surrounded by large open squares are PHL 909 (left) 
and IRAS 14026+4341 (right);
these objects are discussed in \S\ref{phl909} and \S\ref{lobal}, 
respectively.   The star indicates the
position of Mrk\,509. The QSO and ULIG loci, both empirically determined,
are indicated by the dotted rectangles.   \label{sampleplot}}
\end{figure}

The far infrared (FIR) color--color diagram has been used as a tool to
detect and discriminate between different types of activity in the nuclear and
circumnuclear regions of galaxies.   The spectral indices plotted in
the diagram are $\alpha(100, 60)$ and $\alpha(60, 25)$, which are obtained
from the {\it IRAS} flux densities at 100 $\mu$m, 60 $\mu$m, and 25 $\mu$m.
Different kinds of objects, such as QSOs/Seyferts, starbursts, and powerful IR 
galaxies, occupy fairly well defined 
regions in the diagram \citep[see, \eg][]{neu85,deG87,tan88,lip94}.
Objects whose FIR fluxes are dominated by reradiation from dust will
be preferentially located around the black-body region of the diagram,
while those with strong non-thermal emission, such
as optically selected QSOs, are found near the power-law region (see
Fig.~\ref{sampleplot}).

\citet{neu85} plot in the FIR diagram the 22 QSOs detected 
in the {\it IRAS} all-sky survey which have good quality fluxes at
the relevant bands (see their Fig. 1), and note that the majority (19) lie 
close to the line
of constant spectral index (dubbed Power-Law in Fig.~\ref{sampleplot}).
\citet{sto91} noticed that 
the other three QSOs, which fell near the region of the diagram 
occupied by luminous IR galaxies (Mrk\,231, Mrk\,1014, 
and 3C\,48) were also at the time the three QSOs that were arguably 
the three best examples of QSOs that showed apparent stellar tidal tails.
Stockton \& Ridgway proposed that these QSOs are either transitional
objects in an evolutionary scenario, or exceptional or extreme
examples which are distinct from the general QSO population.

\citet{lip94} found an independent correlation using the FIR diagram:
infrared-loud AGN having extreme \ion{Fe}{2} emission have positions 
intermediate
between the black-body and power-law regions, while those having less extreme,
but still strong, \ion{Fe}{2} emission lie closer to the power-law region.
L\'{\i}pari proposed that the strength of the \ion{Fe}{2} emission lines might
be associated with the phase of the starburst/superwind of Type II supernovae
\citep{ter92}.  Extreme \ion{Fe}{2} emitters would then be ULIGs at 
the end phase of a strong starburst, and strong emitters would be 
post-starburst objects.

The FIR diagram provides us with an opportunity to select a complete
sample from those objects which have firm detections in the three
relevant IRAS bands.
The precise color limits we choose (see below) are somewhat arbitrary
and are simply meant to define a reasonable sample of objects spanning a
region in the FIR two-color diagram intermediate between the ULIG and
classical QSO populations.

\subsection{Selection Criteria}\label{selcrit}

Our goal is to collect as complete a sample as possible of low-redshift, 
bona-fide QSOs with an intermediate position in the FIR color---color diagram.
Our selection criteria are:

\begin{enumerate}
\item a luminosity above the cutoff defined for quasars by 
   Schmidt \& Green (1983), \ie\ $M_{B} = -21.5 + 5\,{\rm log}\,h$ 
   (or $M_{B} = -22.1$ for $H_{0} = 75$ km\,s$^{-1}$ Mpc$^{-1}$), 
\item a redshift $z \leq 0.4$,   
\item a declination $\delta \geq -30\degr $,
\item firm {\it IRAS} detections at 25 $\mu$m, 60 $\mu$m, and 100 $\mu$m, and
\item a position in the FIR color--color diagram which is intermediate 
         between the ULIG and QSO loci, \ie\ in the region delimited by
         $-0.8 < \alpha(100, 60) < 0.4$ and $-2.0 < \alpha(60, 25) < -0.8$ 
\end{enumerate}

Our starting point was the sample of 179 QSOs observed with {\it IRAS} 
published by \citet{neu86}.
To this sample we added the
sample of warm objects by \citet{low88}, where they
find 6 previously unidentified QSOs from a list of 187 sources.   Finally,
we added the sample of 91 ULIGs (of which 65 were newly discovered) by 
\citet{cle96a}, since some of these objects have 
AGNs which could possibly have QSO luminosities.   Since none of these
samples is complete, our sample cannot be complete.
However, to the best of our knowledge, we have included {\it every known 
bona-fide QSO} which satisfies the 5 selection criteria.

There are a few objects that will be
either within or outside the sample color limits depending on which published 
{\it IRAS} flux 
values are used.  We give priority to the flux values given by 
\citet{neu86}, since the majority of the QSOs in that sample
have ``pointed observations'' in addition to the all-sky survey observations.
The additional observations result in an increase in sensitivity by a
factor of $\sim5$.   Evidently, no pointed observations are available for
those objects which were discovered by {\it IRAS} itself; we use flux values 
from the Faint Source Catalog for these.

The restriction that objects in the sample must have firm detections at the 
three relevant {\it IRAS} bands (see selection criterion number 4 above) will 
bias the sample towards objects of higher IR luminosity.   Since many QSOs 
have pointed observations that have various detection limits, it is difficult 
to establish the luminosity limit that the selection criterion imposes on the 
sample.   However, by inspecting the IR luminosities, $L_{ir}$, of the QSOs
detected at these three bands, we find that objects out to $z\sim0.1$ have
a mean log $L_{ir} \sim 11.5$, and those with $0.1 \lesssim z \lesssim 0.2$
have log $L_{ir} \sim 11.7$ (a precise definition of $L_{ir}$ is given in
\S\ref{agesum}).
Above redshifts $z \sim 0.25$, only ultraluminous (log $L_{ir} \geq 12$) 
objects are detected at the three bands.

We use the $B$ magnitudes published by \citet{ver96}
as a starting point to select objects.  However, these magnitudes
are not homogeneous and have varying uncertainties which could, in some
cases, push objects in or out of the sample.   Therefore,
in order to have as complete a sample as possible, we have
considered all objects which satisfy every one of the other four selection 
criteria, and which have absolute $B$ magnitudes up to one magnitude fainter 
than the cutoff magnitude.   We then searched the literature for more
accurate $B$ magnitudes $m_{B}$ and calculated new values of $M_{B}$ 
(we assume $H_{0}=75$ km s$^{-1}$ Mpc$^{-1}$ and $q_{0}=0.5$ throughout
this paper) using $M_{B} = m_{B} + 5 - 5$ log$D - k - A_{B}$,
where $D$ is the luminosity distance, $k = - 2.5$ log $(1+z)^{1-\alpha}$ is 
the $k$ correction (where we assume $\alpha=0.3$), and $A_{B}$ is the
Galactic extinction, as given by \citet{sch98}.

With the corrected values of $M_{B}$, there are only two objects 
in the pool that are fainter than, but within one magnitude of, the cutoff
magnitude ($M_{B} = -22.1$) for our sample:  
Mrk\,509, which is highly variable, but has an estimated 
$M_{B} = -21.7$ ($B$ magnitude from \citealt{car96}) 
and Mrk\,231, also with $M_{B} = -21.7$ ($B$ magnitude from \citealt{sur00}).
There is much evidence 
indicating that the nucleus of Mrk\,231 is heavily reddened, with more 
than 2 magnitudes of extinction in B (CS00b and 
references therein).   In contrast, no evidence has been
found for nuclear extinction in Mrk\,509.  
In particular, in a spectropolarimetric study of Seyfert 1 galaxies, 
\citet{goo94} find that there is ample evidence that the 
nuclear light of Mrk\,231 is reddened and that the inferred scattering 
is from dust, while the weaker polarization in Mrk\,509 is more likely 
due to electron scattering, and there is no evidence for nuclear reddening.
Therefore we have chosen to include Mrk\,231 in our sample since,
except for the reddening, it would squarely fit in our sample.
However, as Mrk\,231 does not formally qualify as a bona fide QSO, it may be 
appropriate to leave it out in some analyses of the sample as a whole.

\begin{deluxetable}{lcccrrr}
\tablewidth{6.0in}
\tablecaption{Transition Objects\label{sampletab}}
\tablehead{\colhead{Object Name} & \colhead{IAU}
& \colhead{Redshift\tablenotemark{a}} & \colhead{$M_{B}$\tablenotemark{b}}
& \colhead{25 $\mu$m\tablenotemark{c}} & \colhead{60 $\mu$m\tablenotemark{c}} &
\colhead{100 $\mu$m\tablenotemark{c}}}
\startdata
IRAS\,00275$-$2859&0027$-$289&0.2792& $-$23.16&173$\pm$45 &690$\pm$55 &734$\pm$147  \\
I\,Zw\,1         & 0050+124  &0.0611& $-$22.62&1097$\pm$20&2293$\pm$17&2959$\pm$51  \\
3C\,48           & 0134+329  &0.3694& $-$24.55&160$\pm$08 &770$\pm$09 &1080$\pm$27  \\
Mrk\,1014        & 0157+001  &0.1634& $-$23.97&520$\pm$58 &2377$\pm$56&2322$\pm$130 \\
IRAS\,04505$-$2958&0450$-$299&0.2863& $-$24.30&189$\pm$19 &650$\pm$52 &765$\pm$122  \\
IRAS\,07598+6508 & 0759+651  &0.1483& $-$24.16&535$\pm$37 &1692$\pm$85&1730$\pm$190 \\
Mrk\,231         & 1254+571  &0.0422& $-$21.67&9184$\pm$11&35257$\pm$16&34229$\pm$47\\
PG\,1543+489     & 1543+489  &0.4009& $-$24.27&126$\pm$18 &348$\pm$26 &485$\pm$79   \\
PG\,1700+518     & 1700+518  &0.2923& $-$25.00&220$\pm$21 &480$\pm$36 &482$\pm$88   \\
\enddata
\tablenotetext{a}{QSO redshifts measured from our spectra, except for
I\,Zw\,1 (Solomon \etal\ 1997), IR\,07598+6508 (Lawrence \etal\ 1988),
and Mrk\,231 (Carilli \etal\ 1998)}
\tablenotetext{b}{Values of $M_{B}$ account for Galactic reddening and
k-corrections. We used $B$ magnitudes taken from V\'{e}ron-Cetty \& V\'{e}ron
(1996) for IRAS\,00275$-$2859 and IRAS\,04505$-$2958; Elvis \etal\ (1994)
for I\,Zw\,1 and 3C\,48; Schmidt \& Green (1983) for Mrk\,1014, PG\,1543+489,
and PG\,1700+518; Surace \& Sanders (2000) for IRAS\,07598+6508 and Mrk\,231.
The cutoff value for QSOs, as defined by Schmidt \& Green (1983), is
$M_{B}=-22.1$ for our chosen cosmology.}
\tablenotetext{c}{$IRAS$ flux density values in mJy}
\end{deluxetable}
The final sample is listed in Table \ref{sampletab}.  
The redshifts listed in this table are as measured from our spectra, with
errors typically being $\pm0.0001$.   We have listed redshifts from the
literature as indicated in the table for three objects whose emission
lines are heavily contaminated by \ion{Fe}{2} emission.
References for $B$ magnitudes are as indicated in the table.  The last
three columns list the {\it IRAS} flux densities at 25 $\mu$m, 60 $\mu$m and 
100 $\mu$m, respectively.   Figure~\ref{sampleplot} shows the position
of these objects in the FIR color--color diagram, indicated by open
circles; the 22 other objects that satisfy selection criteria 1 through 4
are indicated by filled squares.

\section{IRAS\,00275$-$2859, IRAS\,04505$-$2958, PG\,1543+489, and I\,Zw\,1
\label{chap5}}

\subsection{Observations and Data Reduction}

Spectroscopic observations were carried out with the Low-Resolution 
Imaging Spectrometer (LRIS; \citealt{oke95}) on the 
Keck II telescope.  We used a 300 groove mm$^{-1}$ grating blazed at 
5000\,\AA\ with a dispersion of 2.44\,\AA\ pixel$^{-1}$.  The slit was 
1\arcsec\ wide, projecting to $\sim$5 pixels on the Tektronix 
2048$\times$2048 CCD.
We obtained two or three exposures for each slit 
position, dithering along the slit between exposures.
In Table \ref{journal4} we show a complete journal of observations, 
including slit positions, total exposure times, and the dates when the
objects were observed.
\begin{deluxetable}{lclccc}
\tablewidth{6in}
\tablecaption{Journal of Spectroscopic Observations \label{journal4}}
\tablehead{\colhead{} & \colhead{PA} & \colhead{Offset}
& \colhead{Dispersion} & \colhead{Total Int.} & \colhead{} \\
\colhead{Object} & \colhead{(deg)} & \colhead{(arcsec)}
& \colhead{(\AA\ pixel$^{-1}$)} & \colhead{Time (s)}& \colhead{UT Date} }
\startdata
IRAS\,00275$-$2859&\phn325.1 & \phn\phn0.0\phn& 2.44 & 3600 & 98 Sep 13  \\
IRAS\,04505$-$2958& $-$21.0  & \phn\phn1.0 E  & 2.44 & 3600 & 97 Oct 10  \\
PG\,1543+489      & $-$30.0  & \phn\phn4.9 S  & 2.44 & 2400 & 97 Jun 12  \\
I\,Zw\,1      & \phn\phn36.6 &   $\sim 10$ N  & 2.44 & 2400 & 97 Oct 10  \\
\enddata
\end{deluxetable}

The spectra were reduced with IRAF, using standard reduction procedures
(see \eg\ CS00b for details).  For  
IRAS\,04505$-$2958, we subtracted the scattered light from the QSO from
the spectrum of the companion galaxy by reversing the 2 dimensional 
spectrum and subtracting it from the original spectrum (the slit position
was chosen so that there would be no light scattered from the star NE
of the QSO in our slit).   For IRAS\,00275$-$2859, we subtracted a 
version of the QSO spectrum from those of the host galaxy, scaled 
according to the amount of broad line contamination seen in the
spectra.   The same procedure was followed for the regions closer to the
QSO in PG\,1543+489, where we obtained a separate 300 s exposure of the
QSO nucleus.   No correction was needed for the spectra of I\,Zw\,1.

We have described our modeling strategy in detail elsewhere (CS97, SCC98,
CS00a, CS00b).  Briefly, we use \citet{bru96} isochrone synthesis models
with solar metallicity and a \citet{sca86} initial
mass function (IMF).   We have previously (SCC98) found that the ages 
determined for the post-starburst populations are remarkably robust with
respect to the different assumptions about the nature of the older stellar
component.  Here we use a 10-Gyr-old model with an exponentially 
decreasing star formation rate with an e-folding time of 5 Gyr as an
old, underlying population.  To this we add a young starburst model and do a 
$\chi^{2}$ fit to the data to determine the best fitting age and relative
contributions of the two populations.  We have found (CS00a) that such 
age determinations typically have a precision of about $\pm50$\%.

We obtained color maps of these objects prior to 
spectroscopic observations in order to choose slit positions.   
All imaging observations at optical 
wavelengths were done with the University of Hawaii 2.2 m telescope 
at f/10 using an Orbit CCD, which yielded an image scale of 0\farcs138
pix$^{-1}$.   For IRAS\,00275$-$2859 and IRAS\,04505$-$2958, we used
a $U^{*}$ filter centered at 3874 \AA\ with a bandpass of 593 \AA\ to
avoid strong emission lines, and we obtained six 1200 s exposures for
each object.   We also obtained four 900 s exposures and four 300 s
exposures for each object, respectively, through a $V$ filter, which
included H$\gamma$ through H$\epsilon$ emission, but avoided the stronger 
H$\beta$, [\ion{O}{2}] and [\ion{O}{3}] emission lines.
We obtained four 900 s $B$-band and three 600 s $V$-band images of 
PG\,1543+489.
Finally, we obtained five 1200 s exposures of I\,Zw\,1 through a $U'$
filter (centered at 3410 \AA\ with a bandpass of 320 \AA ), and six
300 s exposures through a $B'$ filter (centered at 4648 \AA\ with 
a bandpass of 830 \AA), where both filters were chosen, once again, to 
avoid contamination from strong emission lines.  All images were 
obtained in subarcsecond seeing and near photometric conditions, except for
those of I\,Zw\,1, which were taken through cirrus.

Additionally, we obtained high-resolution near-IR images using the 
tip-tilt system at the f/31 focus of the University of Hawaii 2.2~m
telescope.   The $1024\times1024$ QUIRC (HgCdTe) infrared camera 
\citep{hod96} gave a plate scale of 
0\farcs0608 pixel$^{-1}$.   We obtained a $16\times300$~s $H$-band image
of IRAS\,00275$-$2859, a $21\times300$~s $H$-band image
of IRAS\,04505$-$2958, a $10\times600$~s $K'$ image of PG\,1543+489, and a
$76\times60$~s $H$-band image of I\,Zw\,1.  Optical and near-IR images
were reduced with IRAF using standard procedures.

{\it HST} WFPC2 images of IRAS\,00275$-$2859 
and IRAS\,04505$-$2958 were obtained from the {\it HST} data archive.   For 
IRAS\,00275$-$2859 we used two Planetary Camera (PC1) 300~s exposures in 
the F606W filter, and two Wide Field Camera (WF3) 400~s exposures in the 
F814W filter.   For IRAS\,04505$-$2958 we used one 400~s and two 600~s 
PC1 images in the F702W filter.    We removed cosmic rays and combined
images as described elsewhere (\eg\ CS00b).

For several objects, we estimate crude dynamical ages from tidal tail lengths.
Although in one or two cases it might have been possible to estimate dynamical
ages from other criteria, the tidal tail estimate can be applied to most of
our sample, and we prefer to use the same criterion for consistency.

The projected physical length subtended by 1\arcsec\ is 3.5 kpc for 
both IRAS\,00275$-$2859 and
IRAS\,04505$-$2958, 4.3 kpc for PG\,1543+489, and 1.1 kpc for I\,Zw\,1.

\subsection{IRAS\,00275$-$2859}

IRAS\,00275$-$2859 (shown in Fig.~\ref{ir0027}) was the second previously 
unidentified QSO discovered in the {\it IRAS} database \citep{vs87}. 
The spectrum of the QSO (Fig.~\ref{ir0027spec}, bottom panel) shows broad 
Balmer emission lines at $z_{\rm QSO}=0.2792$, no narrow emission lines except
for very faint [\ion{O}{2}] $\lambda3727$, and is otherwise dominated by
strong \ion{Fe}{2} emission.
\begin{figure}[p]
\vspace{2.5in}
\caption
{Space- and ground-based images of IRAS\,00275$-$2859. The $HST$ PC1 F606W 
image (top left) shows diffuse clumps NW of the QSO; a spectrum of 
this area is shown in Fig.~\ref{ir0027spec}.  The WFC3 F814W image 
(top right) highlights the curved extension on the NW.  The 
high-contrast $HST$ WFC3 (bottom left) and ground-based $U^{*}\!-\!V$
(bottom right) images have been smoothed to show the low surface brightness 
features.  In $U^{*}\!-\!V$ map, darker regions indicate more negative values
of $U^{*}\!-\!V$.   The 4\arcsec\ scale bars (equivalent to 14 kpc) shown in 
the left-hand panels also apply to the corresponding panels to the right.
In this and all subsequent images, north is
up and east to the left.  \label{ir0027}}
\end{figure}

IRAS\,00275$-$2859 has been described as having a luminous host galaxy with 
a ``clear tail'' extending to the north-east and ``two nuclei'' separated by
$\sim2\arcsec$, the fainter of which has been interpreted as the nucleus
of a strongly interacting galaxy.   Thus, it has been concluded that this
object is in a pre-merger stage \citep{vad87,hut88b,cle96b,zhe99}.

In the top panels of Fig.~\ref{ir0027} we show an {\it HST} PC1 image of 
IRAS\,00275$-$2859.
It is evident from this low contrast image that the ``second nucleus'' is
rather a bright knot ($M_{R} \sim -17.5$)
2\farcs7 south-east of the QSO nucleus, and that it 
is part of a linear structure in a highly disturbed host.   This knot is
very bright in our $U^{*}\!-\!V$ color map in the bottom right panel of
Fig.~\ref{ir0027}, and its
spectrum (Fig.~\ref{ir0027spec}, middle panel) shows a blue continuum
\begin{figure}[p]
\plotone{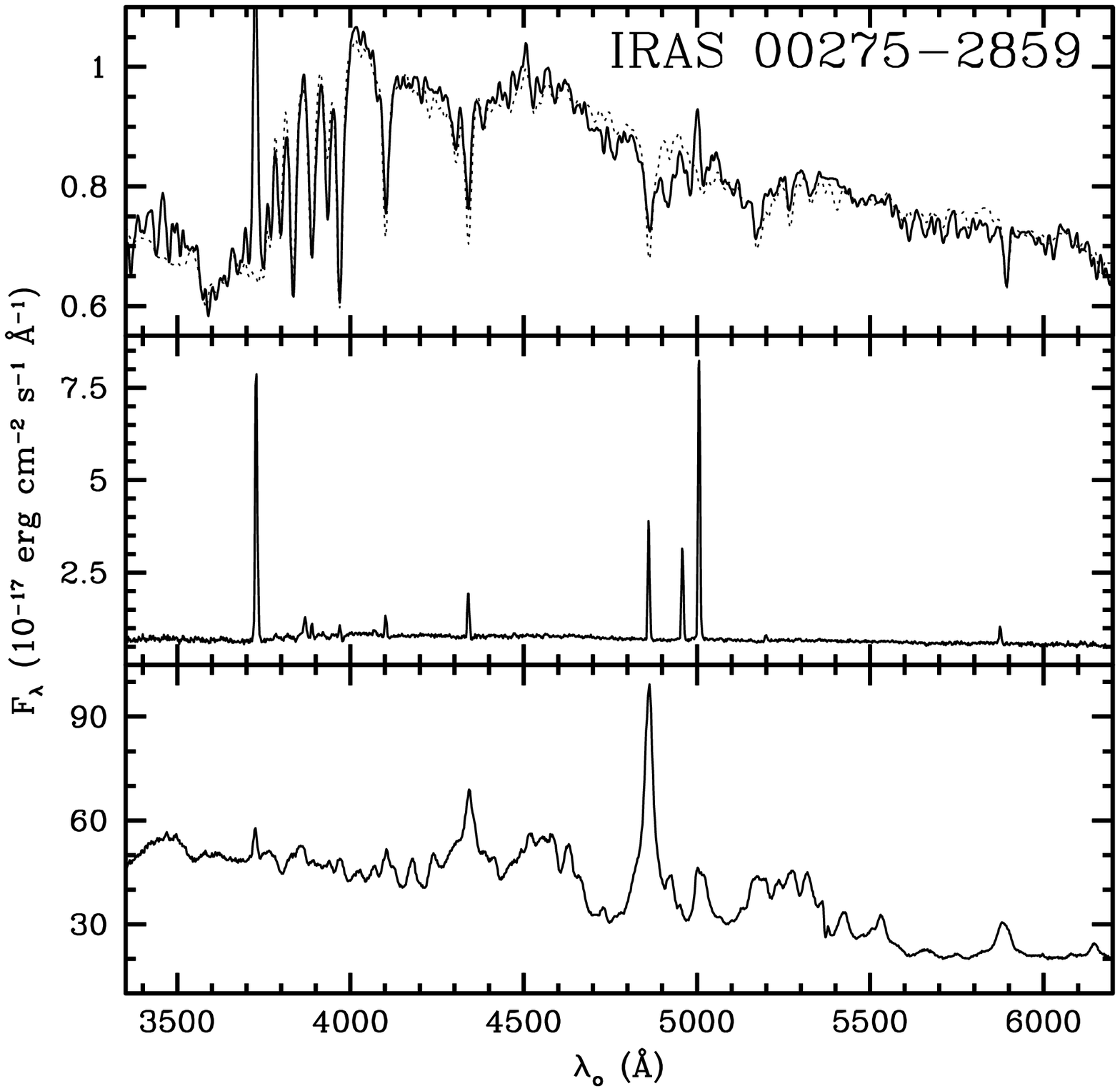}
\caption
{$Top:$ Spectrum of the host galaxy of IRAS\,00275$-$2859.
The dotted line is a model formed by a 50 Myr old instantaneous burst
model and an old underlying population.
$Middle:$ Giant H\,II region in host galaxy $\sim$2\farcs7 southeast
of the QSO nucleus, previously identified as a secondary nucleus.
$Bottom:$ QSO.   All spectra are in rest frame.\label{ir0027spec}}
\end{figure}
with strong emission lines at $z=0.2781 \pm 0.0001$, with a velocity 
gradient of $\sim140$ km s$^{-1}$ from southeast to northwest.   The strong 
emission lines include low 
ionization species such as \ion{He}{1} $\lambda5876$ and [\ion{N}{1}]
$\lambda\lambda5198, 5200$.   Emission line flux ratios indicate that the 
emission is most likely due to gas ionized by young stars rather than by 
the QSO \citep{vei87}.
Thus, instead of being a 
``second nucleus,'' the knot appears to be a giant ($~600$ pc) 
\ion{H}{2} region, roughly the size of 30 Doradus in the Large Magellanic 
Cloud \citep[\eg][]{mey93}. 
This example illustrates that claims of second or multiple nuclei from
morphological data alone (particularly ground-based imaging) should be taken
with some caution.

The {\it HST} PC1 image shows two or more diffuse clumps 
$\sim1\farcs8$ northwest of the QSO.   The spectrum of this region is shown
in the top panel of Fig.~\ref{ir0027spec}; its redshift, measured from
stellar absorption lines, is $z=0.2804$.  
The starburst age obtained from
modeling this region is 50 (+30, $-$20) Myr; the flux contribution from the
old and young populations is roughly equal at rest frame 5000 \AA .   The
spectrum of this region is characteristic of other regions in the host
galaxy, excluding the \ion{H}{2} region.

Our ground-based optical and near-IR images, as well as the {\it HST} WFPC2
images (Fig.~\ref{ir0027}, bottom panels) show clearly the tail on the east 
first reported by \citet{vad87}.  The tail is $\sim12\arcsec$ or $\sim40$
kpc long, and it extends roughly from the position of the \ion{H}{2} region
towards the east and north.   The tail appears redder than the host galaxy
in the $U^{*}\!-\!V$ image (Fig.~\ref{ir0027}), and this may indicate that
the tail is predominantly made of older stars, as suggested by 
\citet{hut88b}.  The tail shows, however, several blue clumps,
possibly regions of star formation like those seen in the tidal tails
of nearby merging galaxies and of 3C\,48 (CS00a) and Mrk\,1014 (CS00b).

A plausible second tail is visible in the {\it HST} WFPC2 image as
well.  This tail extends for $\sim 11\arcsec$ from the northwest of the QSO
(including the clumps described above)  
and arches towards the west and southeast, staying close to the host galaxy, 
and ending in the faint extension southeast of the \ion{H}{2} region.    
This faint extension has a much higher approaching velocity
($-600\pm200$ km s$^{-1}$) than that of the adjacent brighter region closer 
to the QSO ($-190\pm100$ km s$^{-1}$) with respect to the central and 
north regions of the host galaxy.   It appears then that the putative
second tail has a high inclination angle with respect to the plane of the sky,
and that the system has some rotation about an axis perpendicular to the
bright linear feature in the host.   This linear feature, which is also
evident in near-IR images, may indicate that the main body of the host galaxy
is also seen at a high inclination.

Thus, the host galaxy of IRAS\,00275$-$2859 appears to be in the late stages
of a merger of two disk galaxies. 
Assuming a projected velocity of 300 km s$^{-1}$
on the plane of the sky, the tidal tails have dynamical ages of 130 Myr each.
The peak of star formation in the host galaxy seems to have occurred well
after the tails were first launched, as was the case in 3C\,48 (CS00a), 
Mrk\,1014, and IRAS\,07598+6598 (CS00b).

\subsection{IRAS\,04505$-$2958}

The luminous, $z=0.286$, radio-quiet QSO IRAS\,04505$-$2958 was first detected 
by {\it IRAS}, identified as an AGN candidate by \citet{deG87},
and spectroscopically identified as a QSO by \citet{low88}.
Optical images by 
\citet{hut88b} showed a ``double nucleus'' separated by 2\arcsec,
and ``a suggested linear tail along the line of the nuclei, to
the SE''.   Later it was found that the second ``nucleus'', the one to the
northwest, was instead a foreground G star \citep{low89}.
The ``linear tail'' to the southeast, however, was found to be a ``ringlike 
feature'' 1\farcs5 from the nucleus in a {\it HST} F702W PC1 image by 
\citet{boy96} and interpreted as a galaxy that has violently
interacted with the QSO host galaxy.   In Fig.~\ref{ir0450hst} we show a 
lower contrast version of the archival {\it HST} PC1 image, which shows more 
clearly the structure of the ring galaxy.
\begin{figure}[tb!]
\caption
{$HST$ PC1 F701W (left) and $H$-band (right) images of IRAS\,04505$-$2958.
The 4\arcsec\ scalebar is equivalent to 14 kpc.
\label{ir0450hst}}
\end{figure}

\begin{figure}[tb!]
\plotone{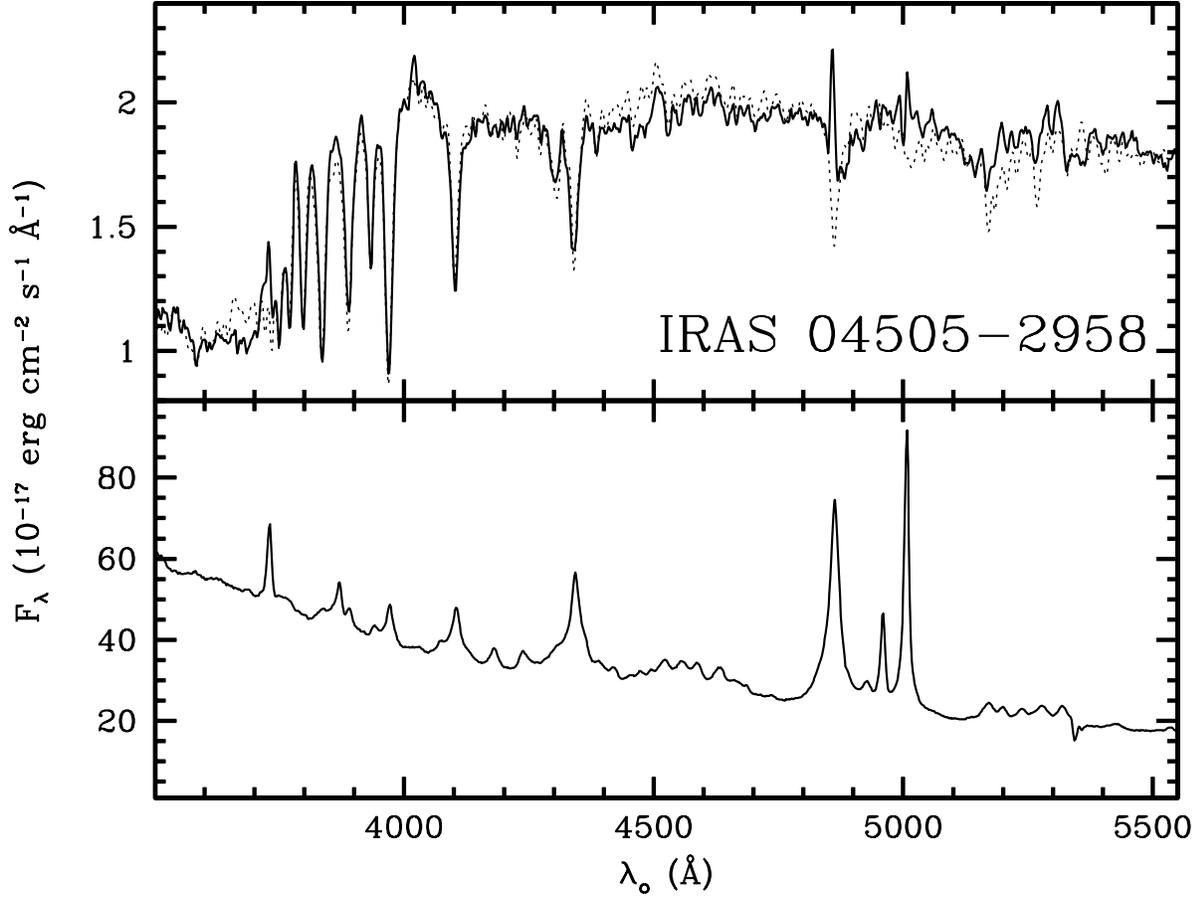}
\caption
{$Top:$ Rest frame spectrum of the companion galaxy to IRAS\,04505$-$2958.
The dotted line is a model formed by the sum of a 128 Myr old starburst
and an old underlying population.  $Bottom:$ Rest frame spectrum of the
QSO.\label{ir0450spec}}
\end{figure}

Figure \ref{ir0450spec} shows the spectrum of the ring galaxy.
The spectrum is dominated by strong Balmer absorption lines; the H$\beta$
and oxygen narrow emission lines most likely come from extended gas ionized
by the QSO and not from star forming regions. 
We obtain an age of 128 (+100, $-$64) Myr for the starburst population,
which contributes $\sim4.5$\% of the total luminous mass along the line of
sight.

It has been found that different individual star forming knots within any
given collisional ring galaxy have virtually no spread in ages
\citep{bra98}, which is to be expected if the star 
formation was triggered by the propagation of a radial density wave at
the time of collision \citep{lyn76}.
We therefore expect the age determined from the integrated spectrum
of the companion to IRAS\,04505$-$2958 to be fairly representative
of the starburst activity in the companion.  

The companion galaxy is clearly visible in our high resolution $H$-band
image (Fig.~\ref{ir0450hst}, right panel) as a slightly elongated feature, 
with a faint
extension to the southwest corresponding to the tidal tail seen in the 
{\it HST} image.  This tail extends for $\sim16$ kpc from the south of
the companion and arching towards the northwest.    The size of the 
brightest part of the ring as measured 
from both the F702W image and the $H$-band image is $\lesssim$4.3 kpc.

The redshift of the ring galaxy is $z_{\rm comp}=0.2865 \pm 0.0002$ as measured
from absorption lines, and that of the QSO is $z_{\rm QSO}=0.2863 \pm 0.0002$
as measured from broad and narrow emission lines.   
The low relative velocity between the QSO and companion may indicate that the 
companion is near apocenter, if the plane of the orbit between the galaxies
is at high inclination (as suggested by the morphology of the companion:
since we see the ring in a fairly open position, and a ring galaxy can be
produced only by a passage with a high inclination to the plane of the
original disk, the mutual orbit is likely to be well out of the plane of
the sky).  Using the tail of the
companion to estimate the time elapsed since the collision, we obtain
$\sim50$ Myr, which is less than the starburst age.   The difference 
between these ages is similar to that of PG\,1700+518, where the dominant
post-starburst population has an age of 85 Myr, but the dynamical age of
the tail of the companion (estimated using similar assumptions) is only 
$\sim40$ Myr.   This lends additional support to the suggestion by 
\citet{hin99b} that these two systems have undergone very similar interactions,
both with small impact parameters.

\subsection{PG\,1543+489 \label{pg1543sec}}

The radio-quiet QSO PG\,1543+489 (Fig.~\ref{pg1543img}) has a redshift 
$z_{\rm QSO} = 0.4009 \pm 0.0001$ as measured from the H$\beta$ and H$\gamma$ 
broad emission lines. The QSO spectrum (bottom panel of Fig.~\ref{pg1543spec})
shows strong \ion{Fe}{2} emission, and no narrow emission lines.
\begin{figure}[tb!]
\caption
{$V$-band and $K'$ images of PG\,1543+489.  Both images have been smoothed
to show the low surface brightness features more clearly.  Scalebar is
43 kpc. \label{pg1543img}}
\end{figure}

\citet{tur97} list PG\,1543+489 as one of 18 
candidate low-redshift BAL QSOs based on the QSO's [\ion{O}{3}] and 
\ion{Fe}{2} emission properties.   They note that while the {\it IUE} spectrum 
of PG\,1543+489 \citep{lan93} has 
very poor signal-to-noise, it appears somewhat absorbed, so that it remains
possible that this object may be a BAL QSO \citep{tur97}.

Figure~\ref{pg1543img} shows the QSO and a compact nearly stellar object
6\arcsec\ (25.3 kpc) south of the QSO.  A bridge between both objects is 
visible in the $V$-band image, but only the smaller clump on the west edge
of the bridge is visible in the $K'$ image.   

The spectrum of the companion (Fig.~\ref{pg1543spec}, top panel) shows 
that it is a galaxy with a stellar continuum and absorption and emission 
lines at a
redshift $z_{\rm comp}=0.4004 \pm 0.0002$, which corresponds to a velocity
difference of $110 \pm 60$ km s$^{-1}$ with respect to the broad line region
in the QSO.   The steep Balmer emission line decrement indicates that the 
companion suffers from dust extinction along the line of sight.
\begin{figure}[p]
\plotone{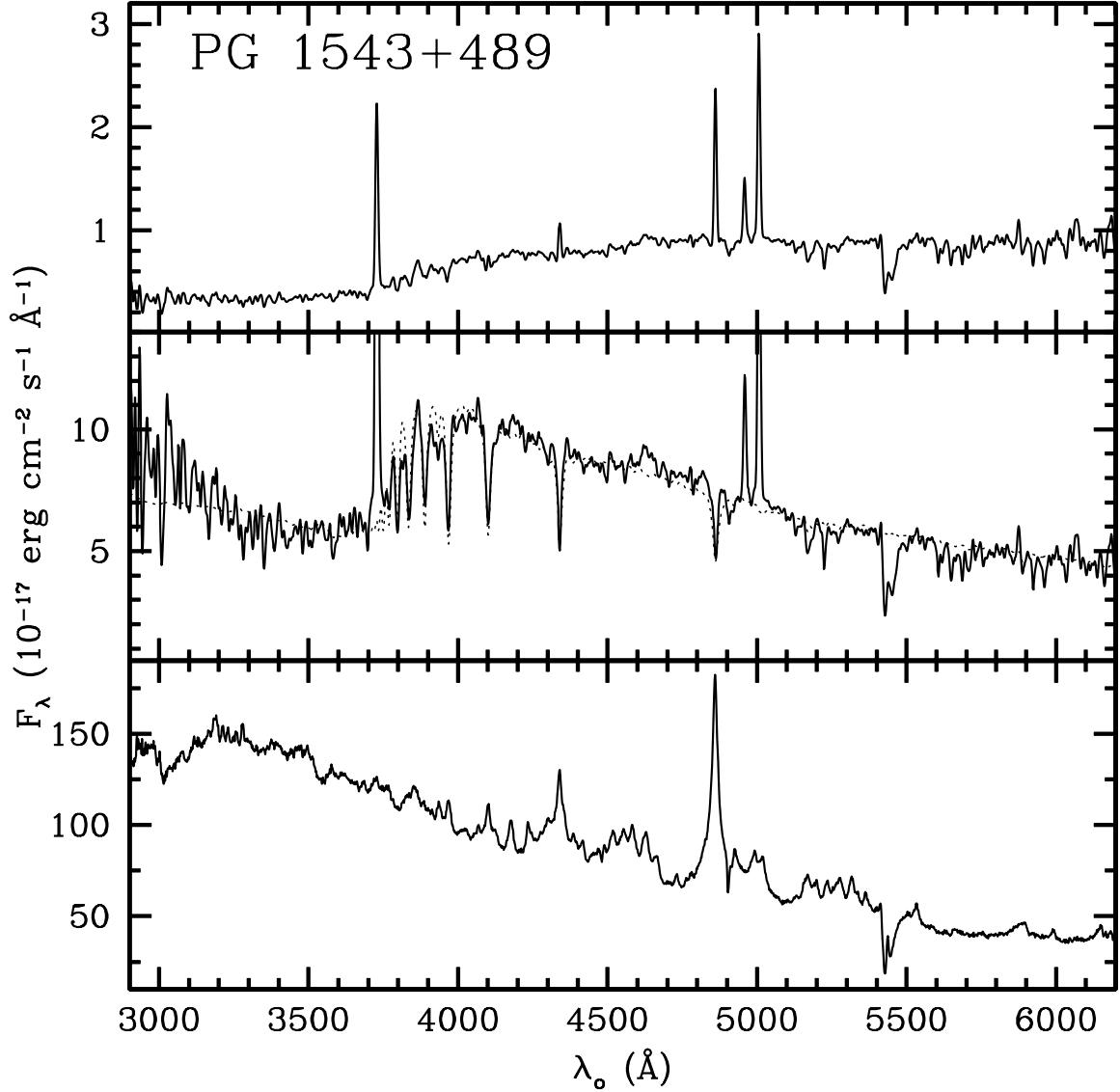}
\caption
{$Top$: Rest frame spectrum of the companion galaxy to PG\,1543+489.
$Middle$: Same spectrum, but dereddened by A$_{\rm v}=1.9$ and with
Balmer emission lines subtracted assuming Case B recombination.
The dotted line is a model formed by a 130 Myr old instantaneous burst
model and an old underlying population.  $Bottom$: Rest frame spectrum of the
QSO. The absorption feature near 5450 \AA in all panels is the atmospheric
B-band.\label{pg1543spec}}
\end{figure}

If we assume that the extinction is caused by a screen of dust located
between us and the companion galaxy, the Balmer line ratios indicate
an extinction A$_{\rm v}\sim1.9$.   Using this very rough estimate, we 
corrected the spectrum of the companion assuming a Galactic 
extinction law \citep{car89}, and subtracted
the Balmer emission lines assuming Case B recombination.   The resulting
spectrum is shown in the middle panel in Fig.~\ref{pg1543spec}.   Modeling
this spectrum with our usual method yields a starburst age of 130 Myr.  This
age is highly uncertain since we have used very rough and simple 
assumptions to correct for reddening, and we also note that the model
clearly diverges from the observed spectrum at short wavelengths.   
However, it is clear that the companion galaxy has a post-starburst spectrum,
and the weakness of the \ion{Ca}{2}, G-band, and \ion{Mg}{1}$b$ features 
indicates that its age cannot be much greater than $\sim$200 Myr.

We have a very faint spectrum of the southwest edge of the host galaxy of
PG\,1543+489.   Unfortunately, the spectrum is too noisy to show 
stellar features.   After correction for scattered QSO light, its
SED is similar to that of the companion galaxy, although it appears
to be slightly bluer.   Faint [\ion{O}{2}] and [\ion{O}{3}] emission
is visible in a small clump along the bridge $\sim2\arcsec$ away from
the companion.

The small projected 
distance and relative velocities between the QSO and the companion, 
the bridge connecting both objects, and the dominant post-starburst
population in the companion galaxy (and possibly in the QSO host) are
all indicative of a strong interaction between these two objects.
The companion galaxy appears as a very compact object ($\leq$ 1.7 kpc) in 
our 0\farcs35 seeing near IR images.   
What we are seeing is perhaps the tidally stripped core of the interacting
galaxy like those proposed by \citet{sto82}.  Even without correcting 
for extinction, the companion has nearly an $L^{*}$ luminosity, and the 
corrected luminosity
is likely to be at least a few $L^{*}$, which is comparable to those
of the companions to PG\,1700+518 \citep{hin99b}
and IRAS 04505$-$2958 \citep{boy96}.
Thus, the companion to PG\,1543+489 and the host
are likely to merge shortly, resulting in a major merger.

\subsection{I\,Zw\,1}

I\,Zw\,1 ($z=0.061$) is near the lower limit of the luminosity 
cutoff for QSOs defined by \citet{sch83},
and it is often classified as a Seyfert 1 galaxy.
The nuclear spectrum of I\,Zw\,1 shows strong \ion{Fe}{2} emission
and weak forbidden-line emission \citep{sar68,bor87}.

The host galaxy of I\,Zw\,1 shows what appear to be two spiral arms, as
well as two ``companion'' objects (Fig.~\ref{izw1img}; see also, \eg\
\citealt{bot84,sur98b,zhe99}).
\begin{figure}[p]
\vspace{3.5in}
\caption
{Top: $U'$ (left) and $B'$ (right) images of I\,Zw\,1.  The $U'$ image has 
been smoothed with 0.5 pixel $\sigma$ gaussian to show the low surface 
brightness features more clearly.   Bottom: $U'\!-\!B'$ color map (left) and 
high contrast $B'$ image (right).  Scale bar shown in upper-left panel
applies to both upper panels and lower-left panel, and is equivalent to
11 kpc.
\label{izw1img}}
\end{figure}
\citet{sto82} reported
that the object to the north is a projected star, while the one to the west 
is a true companion with stellar features.   
The $U'\!-\!B'$ color map shown in the lower-left panel of 
Fig.~\ref{izw1img} shows large, blue
clumps along the arms.   We have obtained spectra of
the companion galaxy and the northwest arm of the host galaxy.

\begin{figure}[p]
\epsscale{1.0}
\plotone{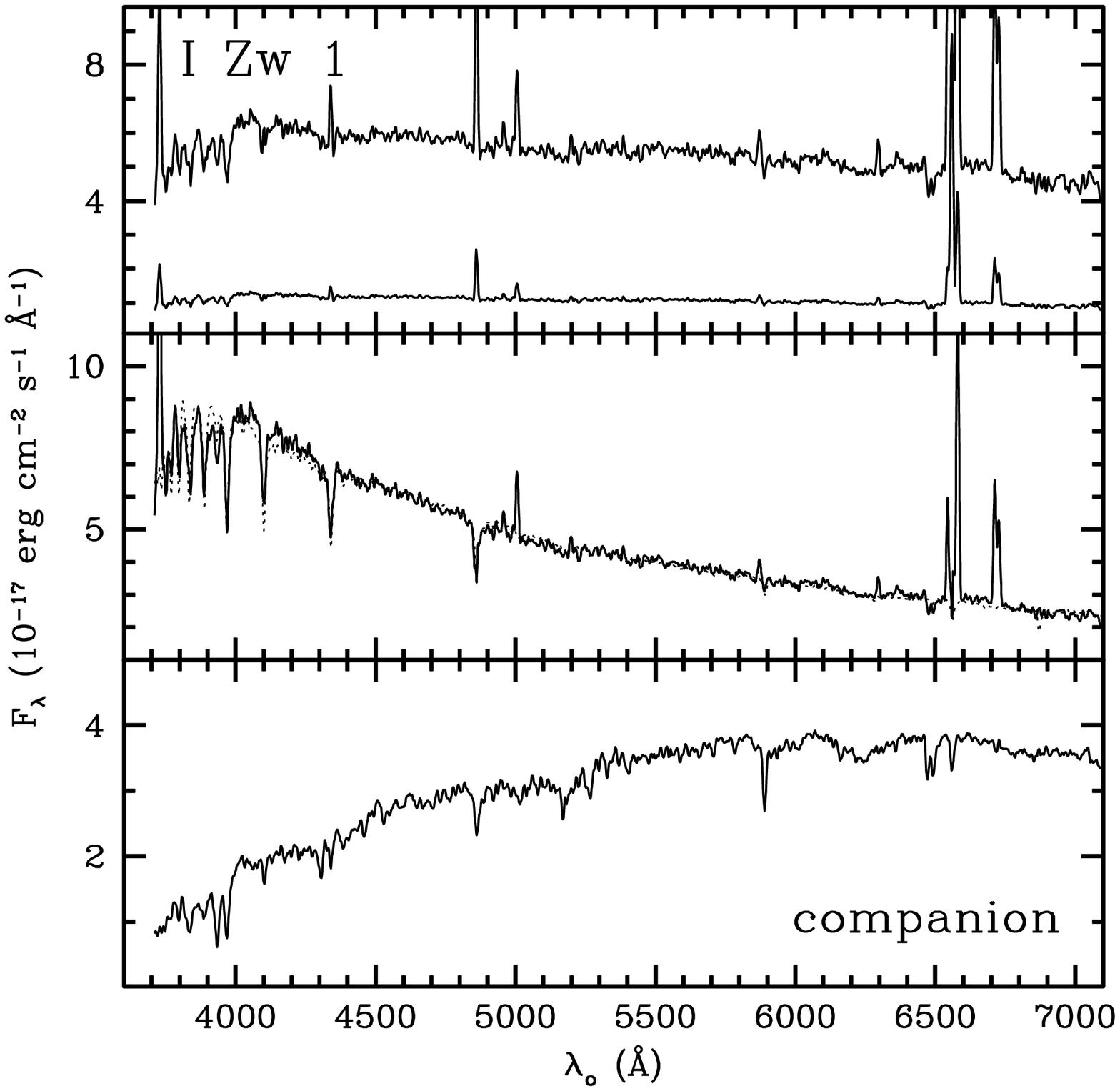}
\caption
{$Top:$ Spectrum of an H\,II region in host galaxy of I\,Zw\,1 (lower trace).
The upper trace displays the same spectrum, scaled by a factor of 5.
$Middle:$ Same spectrum, but corrected for A$_{\rm v}=1.4$ and with
Balmer emission subtracted, assuming Case B recombination.
The age of the starburst, determined from the model (dotted line), is 4 Myr.
$Bottom:$ Companion galaxy to I\,Zw\,1. All spectra are in rest frame.
The absorption feature near 6500 \AA in all panels is the atmospheric
B-band.\label{izw1spec}}
\end{figure}

Fig.~\ref{izw1spec} shows the spectrum of the brightest clump along the
arm, which is characteristic of the spectra in different regions along
the arm (this clump is at a position angle of about $-60$\arcdeg\ from the
nucleus of I\,Zw\,1, close to a point a little less than halfway along a
line joining the companion to the star indicated in Fig.~\ref{izw1img}).   
The spectrum shows strong emission lines
superposed on a stellar continuum.   We measured line flux ratios from 
lines that are close in wavelength (hence fairly insensitive to reddening; 
\citealt{vei87}) and found that they are consistent with 
emission coming from \ion{H}{2} regions.   As in the case of the 
companion to PG\,1543+489,
the steep Balmer decrement indicates extinction from dust along the line of
sight.   Using the assumptions described in \S\ref{pg1543sec}, we corrected
for extinction and subtracted emission lines.   The resulting spectrum is
shown in the middle panel of Fig.~\ref{izw1spec}, where we also plot a
model resulting from a 4 Myr starburst and an old underlying population.
This age is very uncertain, given the assumptions we have made to 
correct the spectrum and the fact that we do not have the blue side of
the spectrum (see section on Mrk\,231 in CS00b).
In any case, this age is consistent with the presence of \ion{H}{2} regions
and, as we have mentioned in CS00a,b, such a young age is simply
indicative of continuous ongoing star formation.

The spectrum of the companion galaxy 15\farcs6 or 16.7 kpc west of the
QSO (Fig.~\ref{izw1spec}, bottom panel)
shows the stellar absorption features of an old population at a redshift 
$z_{\rm comp}=0.0616\pm0.0001$, which, compared with the redshift of the host, 
$z_{\rm host}=0.0610\pm0.0003$, yields a relative velocity of 
$170\pm110$ km s$^{-1}$.   The spectrum does not show any signs of recent
star formation, and it is in fact reasonably fit by a 13 Gyr single burst
population model, assuming there is no significant reddening due to dust.

I\,Zw\,1 shows a seemingly normal disk galaxy harboring a QSO and intense 
star formation, and with strong evidence of a nuclear starburst ring 
\citep{scn98}.  However, the role of a tidal interaction is uncertain.
The morphology of the host galaxy is not indicative of a strong interaction
like those seen in the rest of the objects in our sample.
In addition, the profiles of CO and \ion{H}{1} lines of
the host galaxy are double-horned and symmetric like typical rotating
disks in spiral galaxies \citep{bot84,bar89},
and  VLA 21 cm \ion{H}{1} emission images 
of the host show a nearly unperturbed rotating disk \citep{lim99}.
The companion galaxy, on the other hand, appears to be at least weakly 
interacting with the host.  The low relative velocities indicate that the
companion may be tidally bound to the QSO.  Our deep $B'$ image shows
an extended distribution of very low surface brightness emission, with
the edge of the distribution close to the companion (Fig.~\ref{izw1img},
lower-right panel). Further, H\,I images show an 
extension in the disk of the host both toward and away from the companion 
galaxy \citep{lim99}, which is often regarded as the classic signature
of tidal interactions \citep{too72}.  Our $H$-band image of
the companion (not shown) shows a faint extension away from the host, 
roughly in the same direction that the H\,I disk is elongated (see Fig.~1 
of \citealt{lim99}).   
Thus, I\,Zw\,1 appears to be interacting with its companion galaxy.   
However, the interaction has not produced a burst of star formation 
in the companion (as was the case for PG\,1700+518, IRAS\,04505$-$2958, and
PG\,1543+489).

One possibility is that I\,Zw\,1 is in the process of accreting the low-mass
companion (\ie\ a minor merger in progress), and that the host has developed 
the two-armed spiral pattern in response to the tidal perturbation, as shown 
in numerical simulations \citep{mih94}.   Simulations
also show that such interactions are also capable of triggering substantial 
star formation when the central gas density becomes very large, shortly after 
the spiral arms form and while the companion is still separate from the host, 
if the host lacks a substantial bulge.

\section{Summary of Properties of the Sample of Transition Objects 
\label{agesum}}

Table~\ref{datable} summarizes the observed and derived properties of all
the objects in the sample.   
\begin{deluxetable}{lccrrrcrcccc}
\tablewidth{6.05in}
\tabletypesize{\scriptsize}
\tablecaption{Summary of Properties of ``Transition'' Objects\label{datable}}
\tablehead{\colhead{}&\colhead{QSO}     &\colhead{log $L_{ir}$}&\colhead{$\nu f_ {60}$}&\multicolumn{2}{c}{Starburst\tablenotemark{c}}&\colhead{Tid} & \colhead{Dyn} &\multicolumn{2}{c}{[O\,III]} &\colhead{Fe\,II}       &\colhead{Low} \\
\colhead{Object Name}&\colhead{Redshift\tablenotemark{a}}&\colhead{/$L_{\sun}$}& \colhead{/$\nu f_{B}$\tablenotemark{b}}        & \colhead{Peak} & \colhead{Range}&\colhead{Int\tablenotemark{d}}  & \colhead{Age\tablenotemark{c}} & \colhead{Nucl.\tablenotemark{e}}&\colhead{Ext.}&\colhead{/H$\beta$\tablenotemark{f}}&\colhead{BAL} \\
\colhead{(1)}&\colhead{(2)}&\colhead{(3)}&\colhead{(4)}&\colhead{(5)}&\colhead{(6)}&\colhead{(7)}&\colhead{(8)}&\colhead{(9)}&\colhead{(10)}&\colhead{(11)}&\colhead{(12)}}
\startdata
I\,Zw\,1         & 0.0611 & 11.87& 1.98& (4)& \nodata   &3&\nodata       & 20& N? & 1.5 & N \\
3C\,48           & 0.3694 & 12.81& 3.85&   9\phn &   0---114  &1& 200\phn& 49& Y & 1.0 & N \\
IR\,07598+6508   & 0.1483 & 12.41& 2.03&  32\phn &    32---70 &1& 160\phn&\phn0& w & 2.6 & Y \\
Mrk\,231         & 0.0422 & 12.50&34.65&  42\phn &   0---360  &1& 110\phn&\phn0& N? & 2.1 & Y \\
IR\,00275$-$2859 & 0.2792 & 12.54& 7.19&  50\phn &0---\phn50  &1& 130\phn&\phn0& N? & 1.2 & ? \\
PG\,1700+518     & 0.2923 & 12.58& 1.00&  85\phn &   \nodata  &2&  40\phn&\phn2& w & 1.4 & Y \\
IR\,04505$-$2958 & 0.2863 & 12.55& 2.50& 128\phn &   \nodata  &2&  50\phn& 42  & w? & 0.8 & ? \\
PG\,1543+489     & 0.4009 & 12.67&1.38&$<$200\phn&   \nodata  &2& (80)   &\phn0& N & 1.2 & ? \\
Mrk\,1014        & 0.1634 & 12.52& 4.13& 240\phn &  180---290 &1& 330\phn& 57  & Y & 1.0 & N \\
\enddata
\tablenotetext{a}{QSO redshifts measured from our spectra, except for
I\,Zw\,1 (Solomon \etal\ 1997), IR\,07598+6508 (Lawrence \etal\ 1988),
and Mrk\,231 (Carilli \etal\ 1998)}
\tablenotetext{b}{$f_{60} = f_{\nu}$(60$\mu$m) and $f_{B} = f_{\nu}$(0.44$\mu$m)}
\tablenotetext{c}{Starburst and dynamical ages in Myr.}
\tablenotetext{d}{Type of tidal interaction. 1: major merger, nuclei have
already merged.  2: Galaxies are still distinct and are likely to result in
a major merger. 3: Host interacting with low-mass companion.}
\tablenotetext{e}{Rest equivalent width in \AA . Values given include
anomalous narrow emission in 3C\,48 (CS00a) and blue wing in Mrk\,1014
extending to $\sim -1500$ km s$^{-1}$.  REW for I\,Zw\,1 from Boroson \& Meyers
(1992)}
\tablenotetext{f}{Values of Fe\,II $\lambda4570$/H$\beta$; see text for
references.}
\end{deluxetable}
Column 1 lists the name of each object, and
column 2 its redshift as in Table \ref{sampletab}.   
Column 3 lists log($L_{\rm ir}$) 
in units of solar bolometric luminosity L$_{\sun}$ $= 3.83\times 10^{33}$ 
erg s$^{-1}$.  We have calculated $L_{\rm ir} = L(8\!-\!1000 \mu\rm m)$ 
according to \citet{san96}, 
using $L_{\rm ir} = 4\pi D_{\rm L} ^{2} F_{\rm ir}$, where $D_{\rm L}$ is the
luminosity distance in cm, and 
$F_{\rm ir} = 1.8 \times 10^{-11}\ (13.48 f_{12} + 5.16 f_{25} + 2.58 f_{60} + f_{100})$ [erg s$^{-1}$ cm$^{-2}$], 
with $f_{12}, f_{25}, f_{60}, f_{100}$ being the $IRAS$ flux densities in Jy
given in Table \ref{sampletab}.
Objects with log $(L_{\rm ir}/L_{\sun}) > 11$ are considered luminous 
infrared
galaxies (LIGs), and those with log $(L_{\rm ir}/L_{\sun}) > 12$,
ultraluminous infrared galaxies (ULIGs).

Column 4 lists the ratio $\nu f_{\nu}(60\mu{\rm m}) / \nu f_{\nu}(B)$,
where $f_{\nu}(B) = 10^{-0.4({B}+48.36)}$ [erg cm$^{-2}$ s$^{-1}$ Hz$^{-1}$]
\citep{sch83}, using the $B$ magnitudes described in \S\ref{selcrit}.
This quantity gives a rough measure of the ratio of the SED at IR to optical
wavelengths.

The starburst peak age given in column 5 is the predominant starburst age
found in the host galaxy or companion (see \S\ref{chap5}, CS97, CS00a, CS00b).
We have arranged the objects in order of increasing peak age.  The age of 
I\,Zw\,1 appears in parenthesis to indicate our uncertainty whether the
star formation found has been induced by tidal interaction as in the other
objects.  Column 6
lists the range of starburst ages found in objects with resolved galaxies.
Column 7 indicates the kind of tidal interaction that each system is 
undergoing, as well as the merging stage for those which are major mergers.
Column 8 gives a rough estimate for a dynamical age, in each case assuming
a projected velocity of 300 km s$^{-1}$, and using the length scales
as described for each object (\S\ref{chap5}, CS97, CS00a, CS00b).  
Because of projection effects and other uncertainties, these
estimates will have typical errors of about a factor of 2.
The dynamical age for PG\,1543+489 appears
in parenthesis since we used the length of the bridge between the QSO and 
the companion as the relevant length scale rather than the size of a tidal 
tail.   

Columns 9 and 10 refer to the [\ion{O}{3}] $\lambda5007$ emission from gas 
ionized by the QSOs.
Column 9 lists the rest-frame equivalent widths (REW) measured in the nuclear 
spectra of the QSOs.  As we do not have a nuclear spectrum of I\,Zw\,1, we 
have taken the REW given by \citet{bor92b}. 
Objects with REW$_{[\rm O\,III]} < 5$ \AA\ are considered weak [\ion{O}{3}] 
QSOs \citep{tur97}. Column 10 
indicates whether each object shows luminous extended emission-line regions. 
A ``w'' indicates that the object shows only a much weaker, relatively less 
extended emission.  The classification
is based on the emission observed in our 2-dimensional long-slit spectra
of the objects, and confirmed by our [\ion{O}{3}]
images of 3C\,48 (CS00a) and IRAS\,07598+6508 (CS00b), and results
published by \citet{sto87}  
for PG\,1700+518, PG\,1543+489, and Mrk\,1014.   Our classification for the 
remaining four objects is uncertain (as indicated by question marks) since 
we do not have imaging information, and our slit positions could have missed 
regions of strong emission.

Column 11 gives the ratio of the flux of the \ion{Fe}{2} complex
$\lambda\lambda$4500--4680, formed by multiplets 37 and 38, to 
the flux of H$\beta$.   This \ion{Fe}{2} complex (usually referred to
as \ion{Fe}{2} $\lambda4570$) is often used in measuring ratios since 
it is the easiest and most reliable blend to measure, and its proximity
in wavelength to H$\beta$ minimizes the effects of reddening and uncertainties
in the 
flux calibration.   Whenever possible, we have listed those ratios found 
in the literature where it is clear that the authors have subtracted the 
\ion{Fe}{2} emission before measuring the H$\beta$ flux.   These values are 
close to the values we obtain by simply deblending the lines in our spectra,
except in the case of I\,Zw\,1, for which we do not have a nuclear spectrum.
The references for the ratios we list are:  \citet{bor92a}
for I\,Zw\,1 and PG\,1700+518; \citet{jol91}
for 3C\,48; \citet{lip94}
for IRAS\,07598+6508, Mrk\,231, and IRAS\,00275-2859.
We have listed ratios measured from our spectra for the remaining three
objects.   Objects with 
\ion{Fe}{2} $\lambda4570$/H$\beta > 1$ are considered strong \ion{Fe}{2}
emitters, while those with ratios $> 2$ are called extreme \ion{Fe}{2}
emitters \citep{lip94}. 

Finally, column 12 indicates whether the QSO spectrum of each object 
shows low-ionization broad absorption lines (BAL).  Objects with question
marks either do not have published UV spectra, or (in the case of PG\,1543+489)
their spectra do not have enough signal-to-noise to clearly indicate
whether the object is a BAL QSO.

As can be seen from Table~\ref{datable}, the sample appears to be very
homogeneous.   The selection criteria ensured that these objects would
have similar far infrared colors.  However, we have found that these objects
share many other properties.    All objects are involved in tidal interactions,
and most of them are the result of major mergers, completed or in progress.   
The stellar populations 
are very similar in the different galaxies, with a very small range of 
interaction-induced starburst ages.   In addition, we find a large incidence
of objects with weak [\ion{O}{3}], BAL spectra, and strong \ion{Fe}{2} 
emission.

In the following section, we will consider each one of these properties 
individually.   For each property, we will provide some background, and 
describe and discuss our results.   In \S\ref{together} we will put all 
things together and propose scenarios that incorporate all of our results;
we will also consider the question of evolution.    

\section{Discussion of Individual Properties\label{properties}}

\subsection{Stellar Populations and Interaction Histories \label{agepops}}

We have found that 8 out of 9 objects in the sample have unambiguous tidal 
interaction-induced post-starburst populations.
Of the seven objects that show tidal tails,
we find knots of star formation along the tails, as predicted by numerical
simulations, in at least four cases, and perhaps six.   We also find enhanced
star forming activity in the leading edges of the tails of two of the
objects.   Regions of star formation are found throughout the host galaxies
and, at least in the 5 cases where we have high spatial resolution, the 
strongest and youngest starbursts are invariably concentrated towards the 
center of the host galaxy, clearly indicating the concentration of material 
towards the nucleus expected in mergers.

As discussed in detail in CS00b, this concentration of material
towards the nucleus is likely to have triggered the central strong starbursts
and the QSO activity roughly simultaneously, but we are unable to observe
these initial central starbursts because they are hidden both by the
bright QSO nucleus and by possible more recent starburst activity.
We have chosen to use a ``peak'' starburst age (column 5 in 
Table~\ref{datable}) as a measure of the age
of the QSO activity.   This peak age is the age at which we find evidence
for strongest starburst activity in a given host galaxy, but this is somewhat 
subjective.   As we have seen,
there is an unavoidable uncertainty in using starburst ages to place
QSOs in an time sequence.
This is aggravated in our sample by the fact that we have such a small
range of starburst (as well as dynamical) ages in the host galaxies.
Take the case of Mrk\,231 (see CS00b):
did the onset of the QSO activity occur $\sim$300 Myr ago, when the first
starbursts were ignited, or was it more recently, less than 5 Myr ago,
the time when the star formation in the large H\,II region to the south and
possibly in the knots near the nucleus started?  If we are to take 
the range of starburst ages in a given host galaxy (column 6 in 
Table~\ref{datable}) as the uncertainty in the age of the QSO activity, then 
this uncertainty is almost as large, in some cases, as the 
difference in ages between the oldest and the youngest of the nine
objects.

While the starburst ages may not accurately pinpoint
the age of the QSO activity, they are fairly representative of it.
Simple arguments can show that the age of the QSO activity can be
neither much younger nor much older than the starburst ages.
As we have seen, the starburst ages are consistent with 
having been induced by the evident tidal interactions.   To say that
the starbursts came much later than the QSO activity is then to say that 
galaxies already hosting QSOs were involved in major
mergers that resulted in unusual levels of star forming activity.
While this may very well be possible, it is certainly {\it not probable},
as we will see in detail in \S\ref{chance}.  On the other end, we know that
starbursts did not come much earlier than the QSO activity, since the starburst
ages are already so young.
     
We have 5 confirmed major mergers in our sample, and three additional cases
of strong interactions between galaxies of comparable size which are likely
to merge in the near future.   Only in one case do we have a possible
ongoing minor merger, \ie\ an accretion of a low-mass companion by the host 
galaxy.  Thus, major mergers are the most common triggering mechanism for 
the QSOs in our sample. 
Similarly, the presence of tidal tails in at least 7 of the objects in the
sample implies that disk galaxies are often involved in these mergers.
Whether the parent population of the QSO hosts themselves are usually disk
galaxies is less certain.  Clearly some are, since at least 3 of
our sample show double tails, and I Zw 1 is definitely a spiral.

In terms of merger stage, we have 2 main groups:  (1) objects whose 
nuclei have already merged, and (2) objects where the interacting galaxies 
are still distinct (column 7 of Table \ref{datable}).
By comparing the peak and range of starburst ages to the dynamical age of 
each of these objects, we find a distinction between the two groups:  
except for Mrk\,231,  every object in group 1 has starburst ages that 
are younger than their dynamical ages, while those of group 2 are older 
than their corresponding dynamical ages.
This is consistent with the idea that objects in group 1 had some 
mechanism (possibly a significant bulge in at least one of the galaxies; 
\citealt{mih96}) to stabilize the gas against early dissipation, 
whereas objects in group 2 lacked this mechanism.  The range of ages in
Mrk\,231 may indicate that this object was initially like those of group
2, but that we are seeing it at a time when the galaxies have already
merged.   The fact that we do not see any pre-merger systems (group 2) 
where the tidal tails precede the starburst activity in the hosts suggests 
that the strong peak starburst activity and the ignition of the QSOs in
objects that do have a stabilizing mechanism is delayed until the later
stages of the merger, in agreement with numerical models \citealt{mih96}). 
The fact that we see QSO activity in both groups is further 
evidence that the age of the QSO activity is more closely tied in with the
starburst ages than with the morphology of the merger.
It is clear, however, that the force of these conclusions is limited by 
the size of our sample and by the large uncertainties in the dynamical
ages.

These results unambiguously indicate that we have a population of very young 
objects, both dynamically and in their stellar populations, and, as discussed 
above, with recently triggered QSO activity as well.   We now need to consider
other properties in this context.

\subsection{QSOs and ULIGs: Chance Overlap? \label{chance}}

We have confirmed the long suspected fact that the kind of FIR colors 
in the objects in our sample are indeed indicative of strong star formation 
in the host 
galaxies of QSOs.   We have also shown that the star formation is
not confined to the central regions, but extends at some level over most of
the host galaxies.    These starbursts have been triggered by mergers
rather than by the interaction of some kind of outflow from the QSO with the 
surrounding environments, even in the central regions (at least initially).

ULIGs are defined as galaxies with log$(L_{ir}/L_{\sun}) > 12$, and their most 
common characteristic is that they are mergers: the merger rate in ULIGs is
thought to be at least $\sim 95\%$ (Sanders \& Mirabel 1996 and references 
therein).   
The other common characteristic is the intensified star formation as a result 
of the merger.  So, an object 
that has an infrared luminosity log$(L_{ir}/L_{\sun}) > 12$, widespread 
starbursts, and is a merger, is unambiguously a ``bona fide ULIG''.  
Excluding Mrk\,231 from our sample (on the grounds that its reddened $B$ 
magnitude disqualifies it as a bona fide QSO), we have a sample of 8 
bona fide QSOs, 7 of which are bona fide ULIGs as well (see Table 
\ref{datable}). 

If the QSO and ULIG were two completely unrelated phenomena, what is the 
probability that both would occur in a given galaxy by chance?   In 
order to estimate this probability, we consider the space densities ($\Phi$)
of each of these two types of objects as compared with normal 
galaxies\footnotemark[1].  
\footnotetext[1]{Since we are arguing that QSOs and ULIGs are both rare
when compared to normal galaxies, we make conservative assumptions 
that lead to a lower space density for normal galaxies and higher space
densities for QSOs and ULIGs.  Thus (1) we assume that QSOs and ULIGs
can only occur in galaxies of $L^{*}$ and above;
(2) we use a ULIG luminosity function corrected for the upper end of
our redshift bin, and (3) we use a luminosity function for QSOs consistent
with a luminosity dependent luminosity evolution, which leads to a $\Phi$
a factor of 3 higher than that estimated by the PG survey.}
We transform the different luminosity 
functions (LF) to bolometric luminosities to allow for meaningful comparisons
between the different sets of objects \citep{soi87,san96}.
For normal galaxies, we obtain a $\Phi \approx 3 \times 10^{-3}$ Mpc$^{-3}$ 
by integrating the Schechter LF 
for bolometric luminosities $10 \leq$ log(L$_{bol}$/L$_{\sun}$) $\leq 11.2$.  
The value log$(L_{bol}/L_{\sun})=10$
corresponds roughly to $L^{*}$ if its corresponding $M^{*} _{B} = -19.7$ is
transformed to L$_{bol}$ using the corrections described by \citet{soi86},
and 11.2 is where the LF virtually ends.  For ULIGs, we use the LF given by 
\citet{kim98}, corrected for a redshift $z=0.4$, and assuming a density 
evolution proportional to $(1+z)^{\alpha}$, where $\alpha = 7.6$.  For objects 
with 12 $\leq$ log($L_{bol}/L_{\sun}$) $\leq$ 13, we obtain $\Phi \approx 
1.7 \times 10^{-7}$ Mpc$^{-3}$ (here we have assumed that 
$L_{bol} \approx L_{ir}$; see \citealt{soi87}).  Finally, we use the 
LF for QSOs given by \citet{gra00}
in the range 11.2 $\leq$ log($L_{bol}/L_{\sun}$) $\leq$ 13.3
(roughly equivalent to $-$22.1 $\leq M_{B} \leq -27.6$) and
obtain a $\Phi \approx 2.2 \times 10^{-7}$ Mpc$^{-3}$.   Thus, the fractions 
for ULIGs and QSOs are, respectively, $5.6 \times 10^{-5}$ and 
$7.3 \times 10^{-5}$.   Using these values, the probability that both
phenomena will occur by chance in any given otherwise normal galaxy is 
$4.1 \times 10^{-9}$.

We now consider the expected number of galaxies that show both phenomena
by chance in the integrated comoving volume out to $z=0.4$.   Following
\citet{car92}, the integrated comoving volume, 
$V_{\rm C}$, from the present to a redshift $z$ for an Einstein--de Sitter 
Universe is simply given by
\[ V_{\rm C} = \frac{4\pi}{3} \left( \frac{\rm c}{H_{0}} \int_{0}^{z}
\frac{dz'}{(1+z')^{3/2}} \right)^{3} \; {\rm Mpc}^{3}\]
or $V_{\rm C} = 8\times10^{9}$ Mpc$^{3}$ for $z=0.4$ and $H_{0}=75$.  
We then expect $(3 \times 10^{-3}\ {\rm Mpc}^{-3}) \times (8\times10^{9}\ 
{\rm Mpc}^{3}) \times (4.1 \times 10^{-9}) = 9.8 \times 10^{-2}$ galaxies 
to be both ULIGs and QSOs by chance.   In contrast,
we have at least 7 such objects in this volume.

All of this is to say that the presence of both phenomena in one galaxy
must necessarily be physically related.  By presenting a sample of 8 
bona fide QSOs, 7 of which are also bona fide ULIGs, we are 
firmly establishing the connection between the two phenomena, at least in
the objects in our sample.   Moreover, the commonly held view that QSOs are 
fueled
in strong tidal interactions has up until now been mostly based on 
circumstantial evidence.   Here we have firmly established that at 
least some QSOs can be traced back to a merger and an ultraluminous infrared
phase.

\subsection{Far Infrared Colors \label{firage}}

One of the initial goals of this project
was to put the transition objects on an age sequence.   In an evolutionary
scenario where objects move slowly across the FIR diagram, we would have
expected those objects with younger ages to be closer to the ULIG region,
and those with older ages closer to the QSO region.

\begin{figure}[tb!]
\plottwo{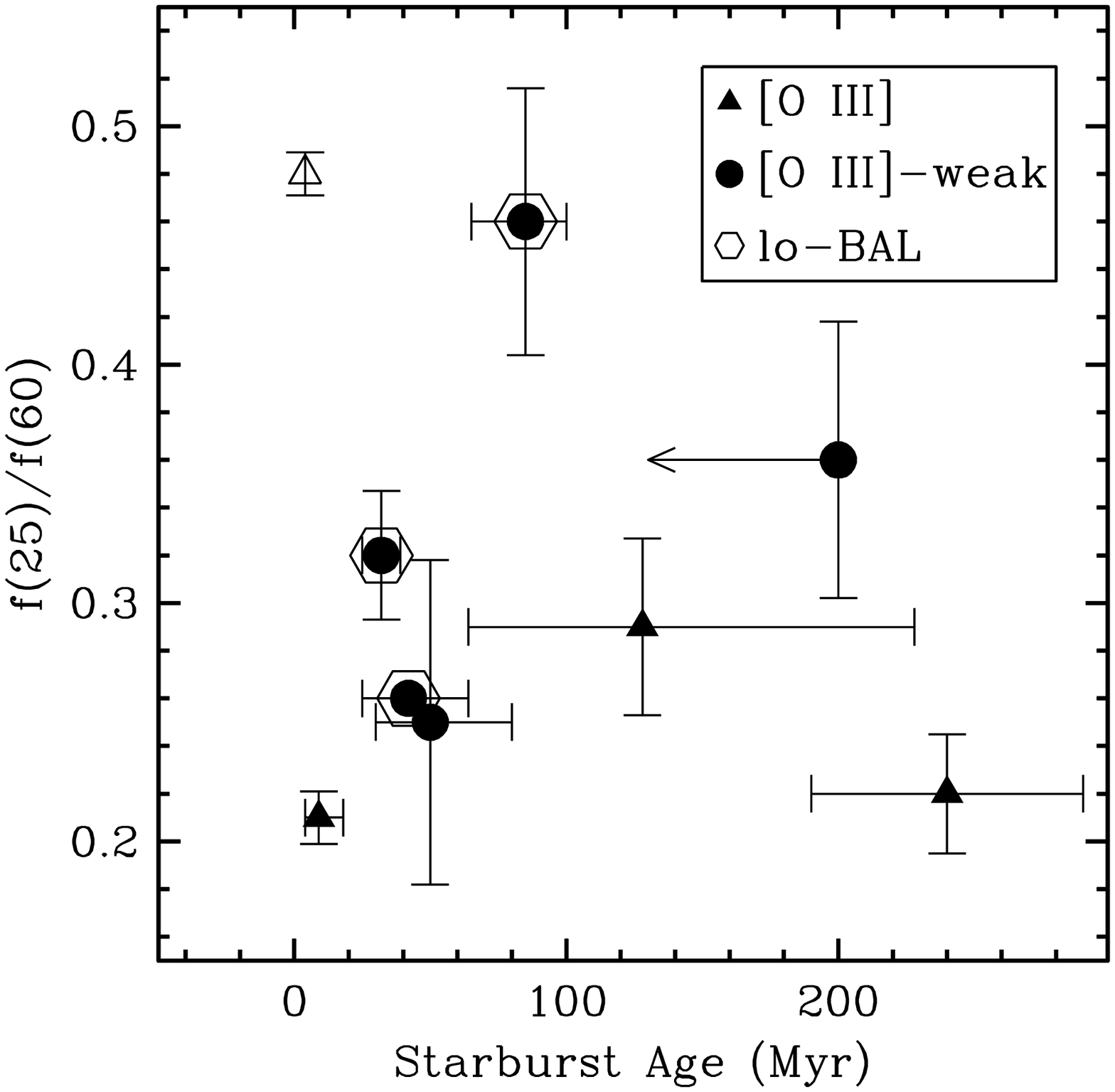}{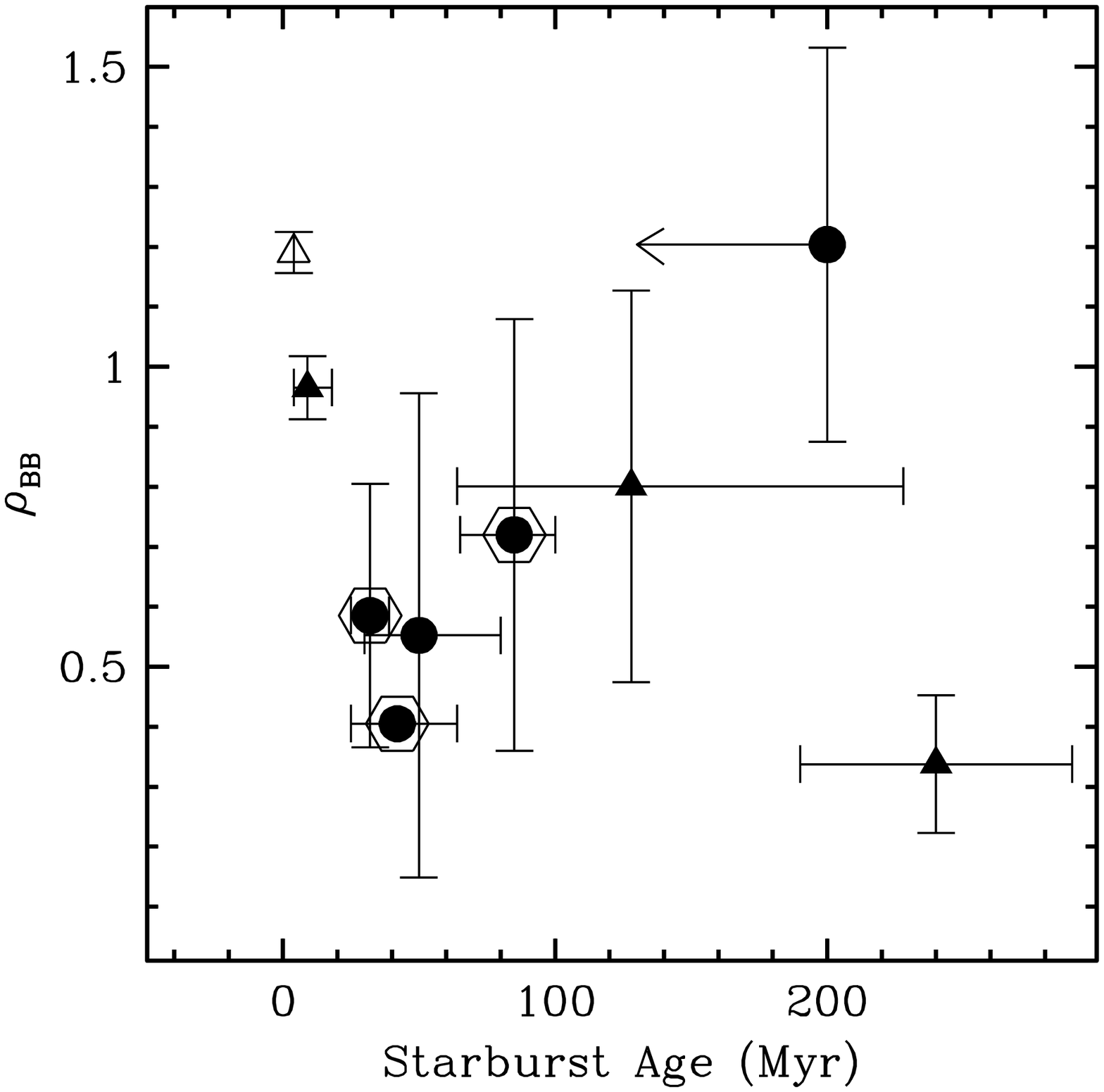}
\caption
{FIR Colors vs. starburst ages.  The vertical axis on the left plot
corresponds to the ratio of $IRAS$ flux densities at 25$\mu$m and 60$\mu$m
while that on the right corresponds to $\rho_{\rm BB}$, the perpendicular
distance from the black body line in Fig.~1.  Objects with
[\ion{O}{3}] emission are plotted as triangles, while those with weak or no
[\ion{O}{3}] are plotted with circles.   I\,Zw\,1 is plotted as an open 
triangle
to indicate the uncertainty in the interpretation of its starburst age (see
text) as well as the fact that it has intermediate-strength [\ion{O}{3}]
emission.  Low-ionization BAL QSOs are indicated by hexagons.
\label{distplot}}
\end{figure}

In Figs.~\ref{distplot}$a$ and $b$ we compare starburst ages for the 
objects with their position in the FIR diagram in two different ways.
First, in Fig.~\ref{distplot}$a$ we plot the ratio of the {\it IRAS}
flux density at 25 $\mu$m to that at 60 $\mu$m, f(25)/f(60), for each
object, against its corresponding starburst age.  The ratio f(25)/f(60)
has been commonly used to isolate candidate objects in searches for 
active galactic nuclei.   
\citet{low88} note that while the 
flux at 60 $\mu$m is in many cases strongly affected by cool emission
from star-forming regions outside the active nucleus, the flux at
25 $\mu$m ``can be even more strongly influenced by activity in the nucleus
and serves as an indicator of either warm dust heated by an intense
radiation field or nonthermal emission.''   The second parameter we
use (Fig.~\ref{distplot}$b$) includes both color indices, and has the form 
$\rho_{\rm BB} \propto y - bx + c$,
where $y$ and $x$ are the indices $\alpha(60,25)$ and $\alpha(100,60)$, 
respectively.  The constants $b$ and $c$ are chosen in such a way that
$\rho_{\rm BB}$ will measure the perpendicular distance to the
line that indicates black-body (thermal) radiation in the FIR diagram.
(Fig.~\ref{sampleplot}).  

It is no surprise that no clear correlation is found in either one
of the plots.   As discussed above, the transition objects have a
rather small range in starburst ages.  Considering the uncertainties in
age determination, in the precise relation between starburst age and time
elapsed since QSO ignition, and in the timescales for buried QSOs to become 
visible (see below),   we do not expect to find a clear correlation between 
age and FIR colors in our sample even for the case where evolution uniquely 
determines the FIR colors of the objects.   

The fact that the ages for all the transition objects are so young and  
span such a small range is, however, an important result in itself.   
The interpretation of this result depends on the relative time scales
of the different physical phenomena in these objects.   Consider two
extreme scenarios:

(1) No evolution in the FIR diagram:  The ULIG phase outlives the QSO 
phase, so that the QSOs in our sample are born in their current
position and spend their entire lives there.   In this scenario, the
small range of starburst ages would indicate that the lifetime of a 
QSO is very short, \ie\ $\lesssim 300$ Myr.

(2) Evolution in the FIR diagram:  The QSO phase is longer than the 
ULIG phase.   The QSOs in our sample were born in the ULIG region of
the diagram, are now in transition, and will eventually end up in 
the classical QSO region.   In this scenario, our results indicate
that the transition stage is very short, \ie\ $\lesssim 300$ Myr.

A third possibility, one where objects move from the classical QSO region
to the transition region (\ie\ galaxies initially hosting QSOs undergo
major mergers and become ULIGs as well) has been dismissed in \S\ref{chance}.
In \S\ref{together} we will discuss whether our results support either one
of the two scenarios proposed above, and in \S\ref{further} we will propose 
additional observations that can help discriminate between the two of them.

\subsection{[O\,III] Emission}

Two points are worth noting regarding [\ion{O}{3}] emission in the 
transition objects.   First, there is a high incidence of [\ion{O}{3}]-weak
objects in our sample (5 out of 9), \ie\ objects whose 
REW$_{[\rm O\,III]} < 5$ \AA .  [\ion{O}{3}]-weak QSOs are uncommon
in samples of radio-quiet QSOs (\eg\ $\sim 10\%$ in \citealt{bor92a}, or
$\sim 20\%$ in \citealt{sto87}),
and even less common in
samples of radio-loud quasars (\eg\ only 1 out of 45 in \citealt{cor97}, or
1 out of 58 in \citealt{bro96}).  
Second, there is a known correlation
between the presence of strong extended emission and strong nuclear 
narrow-line emission in QSOs \citep{bor84,sto87}. 
The two transition objects that show strong extended emission, Mrk\,1014
and 3C\,48, also show the strongest nuclear emission.   We will refer back
to these two points in the following sections.

\subsection{Broad Absorption Line QSOs \label{lobal}}

BAL QSOs are relatively rare, comprising only about 12\% of all radio-quiet
QSOs in current magnitude limited samples.   The broad, blueshifted troughs 
in the spectra of BAL QSOs are 
indicative of high velocity (up to $\sim$ 0.1 c) outflows.   Although BAL 
QSOs may form a 
class independent from non-BAL QSOs, the close similarity in 
emission lines and continuum properties in these two kinds of objects 
would seem to indicate that they are physically the same kind
of object.   Therefore the small fraction of BAL QSOs most likely indicates
either a small covering factor of the absorbing material (so that every
QSO has a BAL region that covers only about 10\% of the solid angle 
subtended at the effective continuum source), or a short-lived phase in
the life of a QSO (see \eg\ \citealt{wey97}  and references therein).

An even rarer class comprising only $\sim 1.5\%$ of all radio-quiet QSOs 
in optically-selected samples is that of low-ionization BAL QSOs (hereafter
lo-BAL QSOs), which show absorption from low-ionization ionic species such as
\ion{Mg}{2}, \ion{Al}{3}, \ion{Si}{2}, and \ion{Na}{1}.  These objects cannot 
generally be explained by simple orientation effects since their properties
show significant departures from those of non-BAL QSOs:  they have stronger
emission from \ion{Fe}{2} and possibly \ion{Fe}{3}, weaker \ion{C}{3}],
Lyman $\alpha$, and \ion{N}{5} emission lines, and are substantially redder
than non-BAL QSOs shortward of 2200 \AA\ \citep{wey91,spr92}.
Thus lo-BAL QSOs are thought to constitute a different class of radio-quiet 
QSOs, having more absorbing material and more dust \citep{voi93,hut98}.

Six out of the nine objects in our sample have published UV spectra with 
enough signal to noise to determine whether they are lo-BAL QSOs 
\citep{lan93,tur97}.
It is remarkable that three of these objects (IRAS\,07598+6508, Mrk\,231,
and PG\,1700+518) are lo-BAL QSOs. 
Given an intrinsic probability of 0.015, the probability that at least 3 
out of the 6 objects with good UV spectra are lo-BAL QSOs by chance is 
$6.5\times10^{-5}$.
However, considering that the nuclear continuum of lo-BAL QSOs is 
reddened, their fraction could be significantly underestimated in
optically-selected samples.   \citet{spr92} estimate
that lo-BAL QSOs may be under-represented in optically selected samples by
about an order of magnitude, so that their fraction is more like 0.15.
This would also at least partially account for the 
high incidence of lo-BAL QSOs in warm {\it IRAS} selected objects
previously noticed by \citet{low89}
and which we also see in our
sample.   Using this corrected fraction, the a priori probability
discussed above raises to 0.05.

We have searched the literature for additional cases of lo-BAL QSOs at
low redshift.  \citet{sow97} have compiled a catalog that includes nearly
every known BAL QSO.  From their catalog, as well as the lists of BAL QSOs
published by \citet{jun91} and \citet{bar97},
we find 19 BAL QSOs with $z < 0.4$ and $\delta > -30\deg$; four of those
are confirmed lo-BAL QSOs, including IRAS\,07598+6508, 
PG\,1700+518, and Mrk\,231.  (The recently discovered class of radio-loud
BAL quasars does not yet include any objects with $z < 0.4$; M. S. Brotherton,
private communication; see also \citealt{bec00}).   Therefore, to the best 
of our knowledge,  our sample contains {\it three of the four confirmed 
lo-BAL QSOs} that fall in our redshift bin and declination range.  
The fourth lo-BAL QSO is IRAS\,14026+4341 ($z=0.3233$, M$_{\rm B}=-24.1$), 
which shows weaker absorptions similar to those of Mrk\,231 (D. C. Hines, 
private communication).   Although not part of our sample, this object 
appears to be closely related to the objects in our sample.  
Fig.~\ref{sampleplot} shows 
that IRAS\,14026+4341 has a position in the FIR diagram 
fairly close to the (relatively arbitrary) limits of our sample.   
Its FIR index $\alpha$(60, 25) is well within the range of our sample,
while $\alpha$(100, 60) is less than 1$\sigma$ from our limits.
In addition, {\it HST} WFPC2 archival images show that this object has an 
extended tail and an interacting companion or 
large starburst region 2\farcs5 (9.6 kpc) away from the QSO nucleus.
Finally, the nuclear spectrum of this object shows no [\ion{O}{3}] emission
\citep{tur97} and strong \ion{Fe}{2} emission (\ion{Fe}{2}/H$\beta
\approx 1.0$; \citealt{lip94}). Thus, it appears that, at least for this
small sample of lo-BAL QSOs, all the objects have very similar FIR and
optical properties.

\citet{bor92b} and \citet{tur97} have noted a
correlation between weak nuclear [\ion{O}{3}] emission (equivalent widths 
less than 5 \AA) and the presence of low-ionization absorption troughs in QSOs.
We have found this tendency in our sample as well.   Of the six objects 
observed in the UV, the three that are lo-BAL QSOs are also [\ion{O}{3}]-weak
QSOs with REW$_{[\rm O\,III]} \leq 2$ \AA , whereas the three objects that are
not BAL QSOs have REW$_{[\rm O\,III]} \geq 20$ \AA\ (see Table~\ref{datable}).
We also confirm the previously noticed tendency for these objects to show
strong \ion{Fe}{2}, which will be discussed further in the following section.

Thus, our observations confirm trends noted before.  To these we add
the following, for at least the small sample of 4 low-redshift lo-BAL QSOs: 
(1) they
have a small range in FIR colors, (2) all four objects have strong
signs of tidal interaction (likely major mergers), (3) at least the three
objects in our sample show interaction-induced enhanced star formation, 
with post-starburst ages $\lesssim 300$ Myr.

Our results then support those interpretations of the lo-BAL phenomenon
which imply young systems, either in the form of young QSOs ``in
the act of casting off their cocoons of gas and dust''
(\citealt{voi93}; see also \citealt{ega96} and \citealt{haz84}),
or as the result of outflows driven by supermassive starbursts 
\citep{lip94,shi96}.  We will return to this point in \S\ref{together}.  

It would be of interest to investigate further the possibility that lo-BAL
QSOs are exclusively associated with young systems by carrying out 
spectroscopic and high resolution imaging observations of the host galaxies
of larger samples of lo-BAL QSOs.
Currently, there are no other studies on the stellar populations
of the host galaxies of lo-BAL QSOs.   In \citet{can98} we have described 
a $z\sim2$ BAL QSO that may be low-ionization (but this needs to be confirmed 
spectroscopically).   The host galaxy shows some asymmetry, and a 
crude estimation of the age of the stellar populations in the host galaxy from 
colors yields $\sim 500$ Myr.   
If we can firmly establish that the lo-BAL phenomenon represents a short
phase in the early life of QSOs and place limits on the duration of this
phase, we could potentially place limits on the mean lifetime of the
QSO activity.   To this end, it would be necessary to also establish what 
fraction of QSOs go through a lo-BAL phase, and an accurate, unbiased value 
for the total fraction of lo-BAL QSOs.  As a numerical example, our current
results would indicate that the mean lifetime of QSOs is about
300 Myr $\times (0.15)^{-1} \times (0.5) = 1$ Gyr,
where 0.15 is the fraction of lo-BAL QSOs (including the correction factor
of \citet{spr92}), and 0.5 is the fraction of the objects in our
sample that undergo a BAL phase (assuming that the homogeneous ages in the
sample indicate that the non-lo-BAL QSOs never went through a lo-BAL phase;
we will see in \S\ref{together} that this may not be the case).   This is,
once again, only an example, as our very small number statistics and the
remaining unknowns prevent us from taking this estimate very seriously.

\subsection{Fe\,II Emission}

Although the \ion{Fe}{2} emission in AGNs has been extensively studied
(\citealt{jol91} and references therein), its origin and nature are still poorly
understood.   The orthodox view is that the  \ion{Fe}{2} emission and other
low-ionization broad lines (such as \ion{Mg}{2} $\lambda$2798) are produced
in high density, high optical-depth regions of broad-line clouds, but current 
photoionization models do not explain the 
large \ion{Fe}{2}/H$\beta$ ratios observed in some AGNs. 

The nuclear spectrum of every object in our sample shows \ion{Fe}{2} emission.
With the exception of IRAS\,04505$-$2958, which has a ratio of 
\ion{Fe}{2} $\lambda4570$/H$\beta = 0.8$, every 
object is either a strong or extreme emitter according to the definition
given in \S\ref{agesum}.   Such high ratios are 
relatively rare in QSOs, with less than 20\% of both radio-loud and 
radio-quiet QSOs being strong emitters 
(\eg\ \citealt{bor92a,jol91,cor97,mci99}).
Only a handful of low-redshift QSOs with extreme \ion{Fe}{2} emission are 
known (L\'{\i}pari 1994), and two of them (IRAS\,07598+6508 and
Mrk\,231) are part of our sample (L\'{\i}pari lists I\,Zw\,1 as an
extreme emitter as well; however, it is unclear how he obtained such a
high \ion{Fe}{2} ratio).   Therefore, it seems likely that the objects in 
our sample have a common characteristic which enables the production of 
strong \ion{Fe}{2} emission.   

\begin{figure}[tb!]
\epsscale{0.5}
\plotone{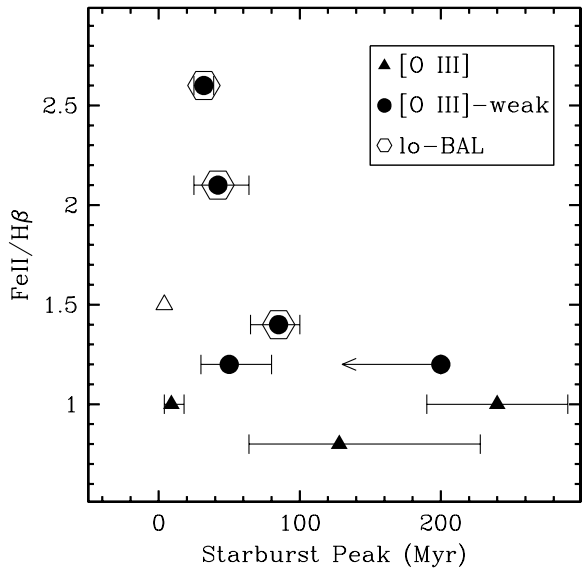}
\caption
{Fe\,II $\lambda4570$/H$\beta$ vs. Starburst Age.  Symbols are as in
Fig.~\ref{distplot}. \label{feiiplot}}
\end{figure}

In \S\ref{samplesec} we described an evolutionary sequence proposed by 
\citet{lip94} connecting the strength of the \ion{Fe}{2} emission with the time
elapsed since a major starburst.    
We find no clear correlation between \ion{Fe}{2} $\lambda4570$/H$\beta$
and starburst peak age (Fig.~\ref{feiiplot}).  Once again, this is not
surprising given the small range and relatively large uncertainties in the 
ages.
While we do not clearly see the trend suggested by L\'{\i}pari, we do find that
the extreme \ion{Fe}{2} emitters are among the youngest objects in our 
sample, and that all of the objects have undergone recent starbursts.
Thus, our results could, in principle, fit with L\'{\i}pari's and Terlevich 
\etal 's (1992) suggestion that the \ion{Fe}{2} emission arises principally 
in superwind generated by the starbursts.   One problem with this model is,
however, that such a
wind would presumably have to produce the velocities characteristic of 
the apparent widths of the \ion{Fe}{2} lines, typically 2000--3000 km s$^{-1}$,
considerably higher than well-documented superwinds.

A different trend which is evident in Fig.\ref{feiiplot} is that 
[\ion{O}{3}] objects (triangles) have weaker \ion{Fe}{2} emission.   
This inverse correlation was initially found by \citet{bor92a}
in low-redshift, radio-quiet QSOs, and it has
been confirmed in samples of both radio-loud and radio-quiet QSOs at
low, intermediate, and high redshift 
\citep[\eg][]{cor96,bro96,mci99}.   Of the 15 strong
\ion{Fe}{2} emitters in Boroson \& Green's sample, 8 have
REW$_{[O\,III]} < 5$ \AA , and all 15 of them have REW$_{[O\,III]} 
\lesssim 20$ \AA\ (with I\,Zw\,1 being at the top of this range).
In our sample, while [\ion{O}{3}] objects (except for I\,Zw\,1) have
the lowest \ion{Fe}{2} ratios, these ratios are still higher
than the median in QSOs ($\sim 0.5$; \citealt{jol91}), and certainly  
higher than other QSOs of comparable [\ion{O}{3}] emission strength.   
Thus, the objects
in our sample are able to produce strong \ion{Fe}{2} emission even
when [\ion{O}{3}] emission is present.

\section{All Things Considered...  \label{together}}

The predominant QSO properties observed in the sample, such as 
strong \ion{Fe}{2}, 
lo-BAL troughs, and weak [\ion{O}{3}] emission, all imply a low ionization 
parameter.  Trends indicate also that, at some level, we have two distinct 
populations: objects with strong [\ion{O}{3}] $\lambda\lambda4959,5007$ 
emission, and those with weak or no [\ion{O}{3}] emission.
There are at least two instances in which QSO spectra can show no
[\ion{O}{3}] emission:
if the upper level of the [\ion{O}{3}] lines is collisionally de-excited, or 
if it is shielded from the 
strong UV continuum.   We now consider a few reasonable models that might 
account for the observed properties of the objects in the sample.

{\bf (1) The broad absorption line region (BALR) is shielding the UV 
radiation:}  
\citet{bor92b}  
point out that the correlation between very weak [\ion{O}{3}] and the presence
of lo-BALs strongly suggests that the BALR has a large covering factor
(close to 1),
preventing much of the ionizing radiation from reaching the low density gas
farther away from the QSO nucleus where the [\ion{O}{3}] emission is produced.
If the BALR has a high physical density as proposed by \citet{bor92b} and
\citet{wey91},
then the excess \ion{Fe}{2} may arise in the BAL clouds themselves 
\citep{spr92}.  This model would be consistent
with the young ages found in our sample if the lo-BAL phenomenon indeed
represents a phase in the early life of QSOs and it would also account for
the observed relation between nuclear and extended~[\ion{O}{3}].   A problem
with this model is, however, that there is a large fraction of objects
having weak [\ion{O}{3}] and strong (or moderately strong) \ion{Fe}{2} that do 
not have BAL spectra, \eg\ only 6 out of 18 [\ion{O}{3}]-weak objects are
BAL QSOs in the study by \citet{tur97},  or only 1 out of 10 in the study 
by \citet{bor92a}. If one takes this model to be the only mechanism 
responsible for producing the rare class of [\ion{O}{3}]-weak objects,
then these numbers imply a small covering factor for the BALR, and this
is at odds with the original premise of this model (\ie\ correlation of 
weak [\ion{O}{3}] and BALs).    A second problem with this
model is that it does not explain the excess \ion{Fe}{2} in the strong 
[\ion{O}{3}] emitters in our sample, \ie\ those objects where the UV radiation
would not be almost completely obscured by the BALR.

{\bf (2) Dusty torus model:}  An obscuring torus of dense, low-ionization
gas and possibly dust surrounding the nucleus was first proposed by
\citet{wey91} to explain the nature of (both high- and low-ionization)
BAL QSOs.  BAL clouds are ablated from the surface of the torus and accelerated
by thermal wind and radiation pressure, and we view objects as BAL QSOs
when the high-velocity outflow lies along the line of sight  (\ie\
when the torus is seen at high inclination relative to the line of sight);
lo-BAL QSOs are then QSOs seen along the most absorbed and dusty lines of sight \citep{bro97}.  
This model has been very successful at explaining the polarimetric properties
of both classes of BAL QSOs \citep[\eg][]{wil92,gle94,coh95,hin95b},
as systems seen near edge-on
are expected to be more highly polarized than those seen face-on.   
Using this model, \citet{wil97} 
have formed an orientation 
sequence for a sample of low-redshift FIR-loud objects which includes
7 of the objects in our sample (with only 3C\,48 and PG\,1543+489 missing
from their sample).   In this sequence, the highly polarized hyperluminous 
infrared galaxies are those QSOs where the torus is seen edge on.  As the
inclination angle decreases, while the polarization is still very high, 
a reddened QSO becomes visible and the BAL clouds (which are the ``skin''
of the torus) come into the line of sight.   At even lower inclinations,
as the torus is seen nearly face-on, the QSO is unobscured and the amount of
polarization drops.  Mrk\,1014 and IRAS\,04505$-$2958 are included in the
latter class, while the other 5 objects in common with our sample are 
included in the
intermediate class (dust torus seen at high inclination). \citet{wil97}
note that this sequence could be explained instead by decreasing dust content,
but argue that the $L_{ir}/L_{opt}$ of the reddened objects does not appear 
to be enhanced with respect to that of the PG QSOs when corrected for
line-of-sight reddening.   The correlation with 
strong \ion{Fe}{2} and weak [\ion{O}{3}] emission is not as obvious in this 
orientation-dependent model.
A possibility suggested by D. C. Hines (\citealt{hin97}, comment
during question period)
is that the \ion{Fe}{2} gas may not be physically associated 
with the torus, but rather lie closer to the central source in dense, 
matter dominated blobs.   While the
side of each blob facing the central engine would be highly ionized, the
side facing away from it and towards the torus would be less ionized and
could give rise to \ion{Fe}{2} emission, which could naturally be seen more
easily in systems closer to edge-on.   In order to explain an orientation
dependent [\ion{O}{3}], however, anisotropic [\ion{O}{3}] emission is required.
\citet{hes93} argue that, unlike the case of the [O\,II] 
$\lambda3727$ line, the relatively higher ionization potential and critical
density of the [\ion{O}{3}] $\lambda5007$ line would allow for a major 
contribution to its integrated luminosity to come from nuclear regions on
a scale not much larger than the broad line region (BLR).   Then the [\ion{O}{3}]
emission could be partially obscured by the dust torus when seen from a 
range of inclinations.  Hes \etal\ point out that if this were the
case, part of the emission may show up weakly in the reflected polarized
spectrum, but this has not been observed in any of the polarimetry studies
of lo-BAL QSOs (see also \citealt{sim98} for an argument against obscured
[\ion{O}{3}] emission).   Even assuming an obscured [\ion{O}{3}] emission, the
correlation between nuclear and extended emission is not explained,
neither is the enhanced \ion{Fe}{2} emission in the three objects
that are supposedly seen near face on.  In addition, the radio properties
of lo-BAL QSOs seem to pose a problem to this model.
\citet{gre00} argue that no preferred 
viewing orientation is necessary to observe BAL systems in the spectra of
quasars since Gregg \etal\ and 
\citet{bec00}  observe BALs in both 
flat and steep spectrum quasars, \ie\ objects that are presumably
viewed along the jet axis and at high inclination respectively (see also 
\citealt{bro97}). 
Finally, the youth in the transition objects would be fortuitous in this model.

{\bf (3) Unusually high densities in the BALR:}   \citet{wey91} propose
an alternate model for lo-BAL QSOs where only the objects falling at the 
very high end of the column-density distribution of BAL QSOs develop 
low-ionization absorption and associated \ion{Fe}{2} emission.  The BALR is 
taken to be hospitable to dust, which would explain the observed FIR emission
and the reddened spectra of these objects.  The covering 
factor is estimated to be 25\%--30\%.
This model then predicts that for every lo-BAL QSO there should be two or three
non-BAL QSOs with comparable IR colors, which is precisely what we see in
our sample (assuming the three objects that have not been observed in the 
UV are not lo-BAL QSOs).  While the weak nuclear [\ion{O}{3}] is not explicitly 
accounted for in this model, one might imagine a scenario where the high 
density in the BAL 
clouds is able to collisionally de-excite the upper level of [\ion{O}{3}].
However, the model does not explain the absence 
of extended [\ion{O}{3}] emission in the majority of the objects in the sample
in spite of the fact that all of them are quite likely to have 
extended gas leftover from the tidal interaction.   In addition, there is no
clear connection between the model and the youth of the objects in our
sample.

{\bf (4) Cocoon-like dust surrounds the nuclear regions of the QSO:}   
We have alluded to several lines of evidence supporting
the presence of large amounts of dust in the nuclear regions of the transition
QSOs (SCC98, CS00a,b, \S\ref{chap5}).   Here we present a model where a 
dust cocoon is shielding the NLR
from the ionizing radiation coming from the central continuum source. 
The resulting low-ionization parameter and the dusty environment would allow 
strong \ion{Fe}{2} emission to arise, while the absence of extended [\ion{O}{3}]
emission would be explained by the inability of the ionizing photons to reach
the extended gas.   The lo-BAL phenomenon would then be interpreted as 
more widespread outflows, and, as proposed by \citet{voi93},
``a manifestation of the [QSO's] efforts to expel a thick
shroud of gas and dust''.   In this model, strong [\ion{O}{3}] objects are more 
evolved, having undergone the lo-BAL phase already, and having thus poked 
some holes through the dusty cocoon.   Ionizing photons are then able to 
reach the NLR through the holes, and also to ionize some of the extended gas.
The idea of an ionizing continuum being able to reach the extended gas only
along some lines of sight is supported by the patchy morphology of the 
extended region in 3C\,48 and Mrk\,1014 (\citealt{sto87}; CS00a,b)
and the fact that much of the gas in their host galaxies is not ionized.
This is also consistent with the fact that the three strong [\ion{O}{3}] objects
are less reddened or obscured than the rest of the objects in the sample.
However, even at this stage, much of the dust is still present in nuclear
regions (see, \eg\ \citealt{kla97} and \citealt{haa98} for dust 
contents in 3C\,48), so that \ion{Fe}{2} emission, while less efficiently than 
in previous stages, is still able to form.
The idea of a cocoon with holes may also explain the different lightpaths in 
BAL QSOs inferred from polarimetric studies, where some continuum is seen to 
escape without passing through dust (\eg\ IRAS\,07598+6508: \citealt{hin95b}). 
A problem with this model is that it is not clear whether a dust cocoon 
alone would be able to completely suppress the [\ion{O}{3}] emission in the
majority of the transition objects while at the same time allowing the broad 
emission lines to be visible, without having to place stringent constraints 
on allowed dust grain sizes.   Other mechanisms which thrive in dusty 
environments may need to be invoked.   For example, \citet{wil96b}
propose an explanation for the strong \ion{Fe}{2}--weak [\ion{O}{3}] correlation
where a dusty low-ionization BLR (more specifically, dense, high-speed, 
\ion{Fe}{2}-emitting clouds) cover the ionizing continuum, thereby 
simultaneously increasing the \ion{Fe}{2} emission and shielding the 
lower-density, more distant gas where the [\ion{O}{3}] emission is produced.
Finally, the youth in 3C\,48 implied by its stellar populations pose a 
problem if 3C\,48 is to be considered a more evolved object.   It is possible
that, being a very powerful object as implied by its very powerful radio 
source, 
3C\,48 was able to break through parts of the dust cocoon more rapidly than 
other objects.   This may, in fact, be the reason why lo-BAL QSOs are seemingly
more rare among radio-loud sources, and so far only one out of 30+ known 
radio-loud BAL quasars has been found to have very strong radio emission 
\citep{gre00}.  

While some of these models may be quite appropriate for classical QSOs,
the results presented in this paper imply that a model describing
transition QSOs must necessarily account for the young ages of these systems.
In principle, each of these models could be modified to account for
the youth in these systems if, for example, the dusty torus described
in (2) was preferentially found in young systems, or the unusually high 
densities of (3) were a direct result of a violent tidal interaction.
For the time being, however, we favor (4) as the working model simply because 
it accounts for all the observed properties and, in particular, it is a
direct consequence of the youth in the systems.

We are now ready to present a cartoon picture of our proposed model:  
Some major mergers between galaxies of similar mass, often disk galaxies, 
trigger intense bursts of star formation. As the gas concentrates in nuclear 
regions, the QSO activity is ignited.  Both the starburst and QSO activity 
start before the galaxies merge if both galaxies lack a substantial bulge, 
or near the
final merger if at least one does have a bulge.  Along with gas, dust is 
concentrated
in the central 1 or 2 kpc, resulting in a dust enshrouded QSO.  Both the
AGN and the starburst heat the dust and the object is observed as a ULIG.
A lo-BAL phase comes next when the QSO casts off its cocoon.
While we are already able to see the QSO, much of the dust cocoon is 
still shielding the ionizing UV radiation, so that \ion{Fe}{2} is able to
survive and oxygen is not doubly ionized in the NLR or extended gas.
As the ionizing photons are able to escape along some lines of sight, 
nuclear and extended [\ion{O}{3}] appear. 
Powerful QSOs, especially powerful radio sources, are able to break through
the dust cocoon more rapidly, and this is the reason why we do not see many 
strong radio-loud quasars in cocoon phase.   
The entire process does not last very long; transition
QSOs either evolve into classical QSOs or die within a few hundred Myr.

In this scenario, our results indicate that we are able to see the
radio-quiet QSOs $\sim 50$ Myr after the peak of the starburst activity.   
This remarkably short time may appear, at first, to be at odds with
the idea that ULIGs evolve to become QSOs, since some ULIGs have been
observed to have starburst ages $\gtrsim$ 300 Myr 
(\citealt{sur98b,tra99}\footnotemark[2]).
\footnotetext[2]{Note, however, that while Tran \etal\ report finding typical 
E+A spectra in these galaxies, they do not include an ``E'' component in
modeling the populations. Therefore their ages should be regarded as upper 
limits when comparing them to the ages we derive.} 
However, on the one hand there must be a dispersion in the properties
of ULIGs that govern the time it takes for dust to be cleared from the
inner regions.   On the other hand, as mentioned in CS00a, and in 
agreement with the scenario 
proposed above, there is almost certainly a range of lines of sight in each 
transition object for which the QSO is hidden and the object is viewed
solely as a ULIG.   In the same way, one might expect that some of the 
objects that we observe solely as ULIGs are seen as QSOs along some lines
of sight.  The overlap in ages is then not only consistent, but also expected
in the proposed scenario.

One might also expect that some of the latter proposed objects (\ie\ hidden 
QSOs residing in ULIGs) have already poked holes not along our line of 
sight and are able to ionize the extended gas.   Therefore, narrow-band
imaging (centered around observed-frame [\ion{O}{3}]) of ULIGs may show 
in some objects
extended emission line regions analogous to those of 3C\,48 or Mrk\,1014
(Canalizo \etal ,  in preparation).
Indeed, the ULIG IRAS\,09104+4109 ($z=0.44$, log$(L_{ir}/L_{\sun})=12.65$), 
which shows a hidden QSO spectrum in polarized, scattered light 
\citep{hin99a},  also shows extended [\ion{O}{3}] emission 
\citep{arm99,kle88}.  \citet{tra00}  suggest that the ionized gas
is tidal debris, as is likely the case in the two transition objects in our 
sample with strong extended [\ion{O}{3}] \citep{sto87}.  
Along similar lines, by considering the properties
that are common to ULIGs that unambiguously show hidden broad line regions
in polarized light, \citet{tra99} find that ULIGs are most likely
to host an obscured QSO if they exhibit a high-ionization spectrum 
characteristic of Seyfert 2 galaxies (see also \citealt{vei97}).
Incidentally, the newly found ``hidden quasar'' presented by 
\citet{tra99} (IRAS\,17345+1124, $z=0.162$, 
log$(L_{ir}/L_{\sun}) \approx 12$) shows optical \ion{Fe}{2} emission.

The eight objects in the sample that are ULIGs have very high IR
luminosities, namely log $(L_{ir}/L_{\sun} \geq 12.4$.  \citet{vei99}
find that $\sim 50\%$ of ULIGs with $L_{ir} \geq 12.3$ show evidence
for nuclear activity, with either Seyfert 1 or Seyfert 2 spectra (the 
latter presumably being hidden QSOs).   This percentage is likely to be 
even higher at log $(L_{ir}/L_{\sun} \geq 12.4$ (D. B. Sanders, private 
communication).   Thus, the objects in the sample may be 
representative of a large fraction of ULIGs at high luminosities, and 
our results could, in principle, be consistent with a scenario where 
a majority of ULIGs at these luminosities go through a transition phase.

It is more difficult to estimate what fraction of optically selected 
QSOs might be represented by our sample.   \citet{san89} show
the distribution of the ratio $L_{IR}/L_{UV}$ for PG QSOs (their Fig.~5),
having a mean ratio of $0.4 \pm 0.15$.  
If we assume that the ratio $\nu f_{60}/\nu f_{B}$ for the sample objects
listed in column 4 of Table~\ref{datable} is representative of the 
ratio given by Sanders \etal\, and we scale it using Mrk\,1014 (which is
a member of both the PG sample and our sample), then the
transition objects in our sample (excluding Mrk\,231, which is not a 
bona fide QSO and is obviously heavily reddened) have a mean ratio of 
$1.28 \pm 0.95$.
By comparing the distributions, we estimate that the transition sample is
representative of $\sim7.5\%$ of the optically selected PG QSO sample.
Taking this number at face value would imply that a considerable fraction 
of optically selected QSOs do not go through a transition phase, unless 
the QSO activity can last as long as $\sim 4$ Gyr.    However, the ratio 
of IR to UV luminosities should be treated with caution.   While the FIR
emission is expected to be emitted isotropically, the flux at UV wavelengths
can be significantly depressed with the presence of a screen of dust along
the line of sight.  Thus this ratio is very likely more indicative of reddening
than of intrinsic characteristics in the objects.

Finally, a word on radio-loud vs.\ radio-quiet QSOs.  Our results do not
shed any light on possible relationships or differences between the two
classes of objects.  However, these results give no indication
that the two classes form differently.   Of nine objects,
we have one radio-loud QSO (3C\,48), which is exactly how many we would expect
in a near-complete sample including both kinds of QSOs.   Of the properties we 
studied in the different objects, all of them were virtually identical
in 3C\,48 and Mrk\,1014: morphologies, starburst timescales, and nuclear
properties.   The newly discovered class of lo-BAL quasars of moderate
radio luminosity \citep{bec00} shows that, in this scenario, 
radio loud objects can indeed undergo an outflow phase.   The fact that 
strong \ion{Fe}{2} emitters are less numerous and [\ion{O}{3}]-weak 
objects are almost non-existent in radio-loud QSOs could be explained
as a selection effect: extended radio structure indicates to some level
that UV radiation has already been able to escape.   This would also explain 
why \ion{Fe}{2} emission appears weaker in lobe-dominant radio-loud QSOs
\citep{wil96a}. Radio-loud QSOs seem to have intrinsically stronger nuclear
and extended [\ion{O}{3}] emission than radio-quiet QSOs 
\citep[\eg][]{bor84,sto87,wil96b},
and this may in turn be indicative of a more powerful
ionizing radiation that can break through obscuring material more rapidly than
radio-quiet QSOs.   In noticing differences between radio-loud and 
radio-quiet QSOs, \citet{wil96a} comments:
``The presence of a hot, dusty, environment with low- and high-ionization (BAL)
outflows along the line-of-sight [in radio-quiet QSOs] therefore has 
something to do with {\it lack} of powerful radio emission''.  In our model,
the connection would simply be that those systems lacking powerful radio 
emission take a longer time to get rid of the very dust that is connected with
such characteristics.

The model we have presented is by no means complete or unique; most 
likely there are elements of each of the different models discussed 
above which play a role in actual systems.   Our intention is simply
to lay the foundations for a working model that can be tested with
observations and, most importantly, that incorporates the new
ingredient we have found in these systems, \ie\ that they are recently
fed AGNs residing in dynamically young systems with young post-starburst 
populations.

And now for the question of evolution.   In \S\ref{firage} we stated that
the young ages found for the sample objects suggest two possible
scenarios:   one where objects evolve in the FIR diagram and the transition
between groups is fast, and the other where objects do not evolve and the
QSO activity is short lived.   After analyzing our results as a whole, we
find no strong evidence that uniquely supports either one of these scenarios.
We may be inclined to favor an evolutionary scenario simply because, based
on the model we propose, transition objects could evolve naturally into
classical QSOs if only given enough time.   However a conclusive 
answer will only be reached as we gain knowledge on the relative timescales 
of the two governing phenomena in these systems.
In the following section, we describe further observations that may help to
settle the question of evolution.

\section{Further Tests on the Evolutionary Paradigm \label{further}}

If we could determine the mean lifetimes for the QSO and ULIG phenomena,
the question of evolution vs. no evolution in the FIR diagram would be,
in principle, unambiguously solved.  Unfortunately, there are no simple
ways to determine these lifetimes.  In \S\ref{lobal} we proposed a method to
estimate the mean lifetime of QSOs from observations of lo-BAL QSOs,
but many unknowns need to be solved for before accurate estimates can
be obtained, even if the method does turn out to be applicable.
For ULIGs, we could possibly observe 
spectroscopically a large sample and set upper limits to the time a galaxy
might spend in this phase.   However, the question of
whether every ULIG harbors an AGN is far from settled, and this alone
could confuse the results.

A different approach to the problem is to try to find the natural descendants 
of the transition objects, \ie\ classical QSOs whose host galaxies contain 
older starburst populations and aging signs of interaction.   
We would select a sample of objects with the same selection criteria
as the sample in this paper, except that the
objects would have a position in the FIR color-color diagram farther
away from ULIGs, in the region labeled QSO.  These objects are marked 
by squares in Fig.~\ref{sampleplot}.

The host galaxies of these objects would then be observed with two main 
objectives:
(a) search for and characterize past interaction or merger activity, and
(b) search for and date aging post-starburst populations.
The presence of merger remnants (or lack thereof) would certainly establish
the role of mergers and interactions in the triggering of the bulk of the 
QSO population.   
The presence of aging starburst populations would imply a starburst-AGN
connection as in the transition objects.   The combination of both results
would conclusively establish the connection between classical
and transition QSOs, and, therefore, ULIGs.    
Finally, results would give a measure for the relative time scales of 
these various phenomena.  

\subsection{Signs of Past Interactions and Mergers
in Classical QSO Host Galaxies \label{remnants}}

Recent {\it HST} imaging of QSO host galaxies indicates that at least
a large fraction of QSOs, both radio-loud and radio-quiet, reside in elliptical
hosts \citep{mcl99,bah97}.
If major mergers indeed play a significant role in the origin of the majority 
of the QSO population, it is possible that these elliptical hosts are rather
merger remnants \citep{too72}.  
If this were the case, one might expect to find fine structure indicative of 
past interaction and ancient merger events in the host galaxies.  This 
tell-tale structure, which includes ripples, 
plumes, tails, and boxy isophotes, is commonly seen in nearby ellipticals 
\citep[\eg][]{mal79,mal80,sch92,ben89},
and indicates that at least some of these early type galaxies were either 
formed or structurally modified by mergers
relatively recently \citep[\eg][]{her92,bar92,her88,her89,sch83,too77}.

In a study of a large sample of nearby E and S0 galaxies, Schweizer and 
collaborators \citep{sch88,sch92,sch90}
used a fine structure index, $\Sigma$, to characterize past merger activity.
The index $\Sigma$ measures four types of fine structure thought to be 
caused by mergers, and therefore is sensitive to the dynamical age
of the merger, and provides a rough measure of the dynamical youth or
rejuvenation of the galaxies.  
\citet{san00} found an anticorrelation
between $\Sigma$ and X-ray excess in early type galaxies, which indicates
that galaxies with high fine-structure content have not had enough time
after the merger to form hot halos.

Ground-based imaging observations using adaptive optics (AO) and space-based 
imaging observations using the {\it HST} highly sensitive Advanced Camera for 
Surveys (ACS) would make it possible to detect such fine structure 
in the hosts of QSOs.   The $\sim 0$\farcs1 resolution of such observations
is equivalent to 0.4 kpc for objects at $z=0.4$, even better than the 0.5
kpc resolution in Schweizer \& Seitzer's study of nearby galaxies.
Thus, it is quite possible to apply these techniques to QSO host galaxies
and characterize past interaction and merger activity in these systems. 

\subsection{Aging Starbursts in Classical QSO Host
Galaxies}

As we have seen, spectroscopic observations allow us to date directly the 
ages of the stellar populations in the host galaxies of QSOs.  We can 
use similar techniques to the ones described in this paper to observe
and model the host galaxies of classical QSOs.

In order to test the feasibility of detecting older starbursts 
in host galaxies, we have carried out simulations of different 
intermediate-age starburst populations superposed 
on the spectrum of an old elliptical galaxy.
The simulations show that it is possible to detect and identify
a starburst population as old as $\sim 1.3$ Gyr that comprises 10\% of
the total luminous mass along the line of sight.  The best age indicators 
are the Balmer lines (H$\beta$ and H$\delta$ in particular), the 4000 \AA\ 
break, the CN-band, and, for higher-z objects, the slope of the continuum 
shortwards of 3100 \AA .
The flux of starburst populations older than $\sim 1.6$ Gyr drops very 
rapidly with age and it becomes difficult to distinguish between a
starburst+elliptical spectrum and the elliptical spectrum itself,
unless the mass fraction of the starburst population along the line of 
sight is significantly higher than 10\%.

These constraints give us much room to find the descendants of
transition objects.  As we have seen, our results
imply that objects move rapidly from the ULIG to classical QSO population
in the evolutionary scenario.   Therefore we expect to find descendants
(if any) with post-starbursts ages ranging from $\gtrsim 300$ Myr to
the upper limit of the QSO lifetime.    Dating such relatively young
starbursts should be quite feasible.

\subsection{An Example:  PHL\,909\label{phl909}}

As a preview to what we might expect from these observations, we briefly
discuss results for one member of the classical QSO population, PHL\,909
($z=0.171$), for which analogous observations already exist. 
The position of PHL\,909 in the FIR diagram is indicated in 
Fig.~\ref{sampleplot} by a filled square surrounded by a large open square.
Fig.~2 of \citet{bah97}
shows a PSF subtracted, {\it HST} WFPC2 image
of PHL\,909 through the F606W filter.   Bahcall \etal\ report that the host 
is a normal E4 galaxy.  \citet{mcl99}
successfully model the host with an elliptical 
template in a F675W {\it HST} image, which presumably avoids strong emission
lines.  The QSO+elliptical-host subtracted image shown in their
Fig.~A11 has structure in the residuals which, pending careful
analysis, may be interpreted as the tell-tale structure described in 
\S\ref{remnants}.   
A 900 s Keck LRIS spectrum of the host galaxy of 
PHL\,909 (Canalizo \& Stockton, in preparation) shows that there
is no very young post-starburst component present.  However, this spectrum
does not have sufficient S/N to constrain the presence of a significant
$\sim1$-Gyr-old population.

\section{Summary and Conclusions\label{summary}}

We have studied a sample of transition QSOs, \ie\ bona fide QSOs that have 
far infrared colors intermediate between ULIGs and QSOs.  In summary,

1. Every transition QSO shows signs of tidal interaction.  
Five out of nine are confirmed major mergers and are in the late stages
when the nuclei have already merged.   Three additional objects have 
strongly interacting companions with masses comparable to those of the hosts,
and are expected to merge soon (resulting in major mergers as well).  Only one
transition QSO is undergoing a minor merger.

2. Disk galaxies are often involved in these mergers.
Seven objects show clear tidal tails, and one shows a bridge connecting
the companion galaxy with the host.  
Of the objects that have not yet merged, two are likely to be ring galaxies 
(with tidal tails as well) resulting from near head-on collisions.

3. Every transition QSO shows strong recent star-forming activity, 
and in eight cases this activity is closely related to the tidal interaction. 
These eight objects have spectra characteristic of E+A galaxies.   
The spectra are successfully modeled
by an underlying old population plus a superposed young 
instantaneous burst population.   We find very young starburst ages
for all the objects in our sample, ranging from 
ongoing star formation to $\sim300$ Myr. 

4. By comparing the starburst ages of the central stellar components
with those of the more extended emission, we have
determined the relative ages between stellar populations in various regions
of the host galaxies.  These estimates, along with dynamical ages, place 
constraints on the
timescale for concentrating material in the nucleus, which in turn could
provide the fuel for the QSO.  This is an important step towards
understanding the fueling of QSOs, and the timescales for their
evolution.

5. Interaction and star-forming activity histories indicate that the age
of the QSO activity is more closely correlated with the age of the peak 
starburst than to the dynamical age of the system.

6. Seven out of the eight bona fide QSOs in the sample are also bona fide
ULIGs.   The probability that both phenomena coexist by chance in any given 
otherwise normal galaxy if the two are unrelated is $4.1\times10^{-9}$.  The 
expected number of galaxies 
that show both phenomena by chance in the integrated comoving volume spanned
by our sample is $\sim0.1$ galaxies.  Our sample firmly establishes the
physical connection between the two phenomena in at least one subclass of QSOs.

7. The young ages derived for the transition QSOs imply one of two
scenarios:   (a) At least some ULIGs evolve to become classical QSOs,
and the transition time between the two is short ($\lesssim 300$ Myr), or
(b) At least some QSOs are born under the same conditions as ULIGs and
their lifetime as QSOs is very short ($\lesssim 300$ Myr).

8. We find a high incidence of lo-BAL QSOs, [\ion{O}{3}]-weak QSOs,
and QSOs with strong or extreme \ion{Fe}{2} emission in our sample.  We 
confirm previously reported trends in these properties, and our results 
require that they be interpreted in the context of young systems.

9. Three out of the four known lo-BAL QSOs with $z\leq0.4$ and
$\delta\geq-30\deg$ are part of the sample of transition QSOs.  The 
fourth object has very similar FIR colors, a clear tail, and possibly
luminous regions of recent star formation.   This indicates that the
lo-BAL phenomenon is directly related to young systems, and it
may represent a short-lived stage in the early life of a large fraction 
of QSOs.

10. We find evidence indicating that objects with strong [\ion{O}{3}] emission
and those with weak or no [\ion{O}{3}] represent two distinct populations.
The former may be at a more evolved stage in an evolutionary scenario.

11. We propose a model involving a dust cocoon that initially 
surrounds the QSO nuclear regions, which can account for all the observed 
and derived properties in transition objects.   This model is at least 
qualitatively consistent with the idea of young, dust-enshrouded QSOs
originally proposed by \citet{san88}  

At the beginning of this paper, we posed two important questions
regarding the transition objects: {\it ``Where did they come from?''}, and 
{\it ``Where will they go?''}   We have 
answered the first question unambiguously:  transition objects are the
product of mergers resulting in vigorous star formation and the ignition
of QSO activity.   Transition objects have come from, and are still
part of, the ULIG population.
We have also narrowed down the answer to the second question to two
possibilities:  either transition objects will indeed go on to the become 
part of the classical QSO population within a few hundred million years, or 
they will die within this same amount of time.  In addition, we have outlined 
observations that can potentially discriminate between the two 
possibilities.   The final answer to this question will also provide crucial
answers regarding the timescales involved in the ULIGs and QSOs.

\acknowledgments
We thank Gerbs Bauer, Barry Rothberg, and Scott Dahm for assisting in some
of the imaging observations.  We thank Dean Hines and Mike Brotherton for 
providing information about BAL QSOs prior to publication, and Dave Sanders,
John Tonry, and Esther Hu for helpful discussions.  We are grateful to the 
referee, Sylvain Veilleux, for his careful reading of the paper and useful 
comments and suggestions.   This research has made 
use of the NASA/IPAC Extragalactic Database (NED) which is operated by the 
Jet Propulsion Laboratory, California Institute of Technology, under contract 
with the National Aeronautics and Space Administration.   
This research was partially supported by NSF under grant AST 95-29078.
Part of this work was performed under the auspices of the
U.S.\ Department of Energy by the University of California Lawrence
Livermore National Laboratory under contract No. W-7405-ENG-48.


\begin{thebibliography}{}
\bibitem[Allen, Roche, \& Norris(1985)]{all85} Allen, D. A., Roche, P. F.,
\& Norris, R. P. 1985, \mnras, 213, 67
\bibitem[Armus, Heckman, \& Miley(1987)]{arm87} Armus, L., Heckman, T., \&
Miley, G. 1987, \aj, 94, 831
\bibitem[Armus, Soifer, \& Neugebauer(1999)]{arm99} Armus, L., Soifer, B. T.,
\& Neugebauer, G. 1999, Ap\&SS, 266, 113
\bibitem[Bahcall et al.(1997)]{bah97} Bahcall, J. N., Kirhakos, S.,
Saxe, D. H., \& Schneider, D. P. 1997, ApJ, 479, 642
\bibitem[Bahcall, Kirhakos, \& Schneider(1995)]{bah95} Bahcall, J. N.,
Kirhakos, S., \& Schneider, D. P. 1995, \apj, 450, 486
\bibitem[Balick \& Heckman(1983)]{bal83} Balick, B. \& Heckman, T. M. 1983,
\apj, 265, 1
\bibitem[Barnes(1992)]{bar92} Barnes, J. E. 1992, \apj, 393, 984
\bibitem[Barvainis, Alloin, \& Antonucci]{bar89} Barvainis, R.,  Alloin, D.,
\& Antonucci, R. 1989, \apjl, 337, L69
\bibitem[Barvainis \& Lonsdale(1997)]{bar97} Barvainis, R. \& Lonsdale, C.
1997, \aj, 113, 144
\bibitem[Becker et al.(2000)]{bec00} Becker, R. H., White, R. L., Gregg, M. D.,
Brotherton, M. S., Laurent-Meuleisen, S. A., \& Arav, N. 2000, \apj, 538, 72
\bibitem[Bender et al.(1989)]{ben89} Bender, R., Surma, P., Dobereiner, S.,
Mollenhoff, C., \& Madejsky, R. 1989, A\&A, 217, 35
\bibitem[Boroson \& Green(1992)]{bor92a} Boroson, T. A., \& Green, R. F. 1992,
\apjs, 80, 109
\bibitem[Boroson \& Meyers(1992)]{bor92b} Boroson, T. A., \& Meyers, K. A.
1992, \apj, 397, 442
\bibitem[Boroson \& Oke(1982)]{bor82a} Boroson, T. A., \& Oke, J. B. 1982,
\nat, 296, 397
\bibitem[Boroson \& Oke(1984)]{bor84} Boroson, T. A., \& Oke, J. B. 1984,
\apj, 281, 535
\bibitem[Boroson \& Oke(1987)]{bor87} Boroson, T. A. \& Oke, J. B. 1987,
\pasp, 99, 809
\bibitem[Boroson, Oke, \& Green(1982)]{bor82b} Boroson, T. A., Oke, J. B., \&
Green, R. F. 1982, \apj, 263, 32
\bibitem[Boroson, Persson, \& Oke(1985)]{bor85} Boroson, T. A., Persson, S. E.,
\& Oke, J. B. 1985, \apj, 293, 120
\bibitem[Bothun et al.(1984)]{bot84} Bothun, G. D., Romanishin, W., Strom, S.
E., \& Strom, K. M. 1984, \aj, 89, 1300
\bibitem[Boyce et al.(1996)]{boy96} Boyce, P. J., Disney, M. J., Blades, J. C., 
Boksenberg, A., Crane, P., Deharveng, J. M., Macchetto, F. D., Mackay, C. D.,
\& Sparks, W. B. 1996, \apj, 473, 760
\bibitem[Boyce et al.(1998)]{boy98} Boyce, P. J., Disney, M. J., Blades, J. C.,
Boksenberg, A., Crane, P., Deharveng, J. M., Macchetto, F. D., Mackay, C. D.,
\& Sparks, W. B. 1998, \mnras, 298, 121 
\bibitem[Bransford et al.(1998)]{bra98} Bransford, M. A., Appleton, P. N.,
Marston, A. P., \& Charmandaris, V. 1998, \aj, 116,~2757
\bibitem[Brotherton(1996)]{bro96} Brotherton, M. S. 1996, \apjs, 102, 1
\bibitem[Brotherton et al.(1997)]{bro97} Brotherton, M. S., Tran, H. D.,
Hien, D., van Breugel, W., Dey, A., \& Antonucci, R. 1997 \apjl, 487, L113
\bibitem[Bruzual \& Charlot(1996)]{bru96} Bruzual A., G. \& Charlot, S. 1996,
unpublished [ftp://gemini.tuc.noao.edu/pub/charlot/bc96]
\bibitem[Canalizo \& Stockton(1997)]{can97} Canalizo, G., \& Stockton, A. 1997,
\apjl, 480, L5 (CS97)
\bibitem[Canalizo \& Stockton(2000a)]{can00a} Canalizo, G., \& Stockton, A. 
2000a, \apj, 528, 201 (CS00a)
\bibitem[Canalizo \& Stockton(2000b)]{can00b} Canalizo, G., \& Stockton, A. 
2000b, \aj, 120, 1750 (CS00b)
\bibitem[Canalizo, Stockton, \& Roth(1998)]{can98} Canalizo, G., Stockton, A.,
\& Roth, K. C. 1998, \aj, 115, 890
\bibitem[Cardelli, Clayton, \& Mathis(1989)]{car89} Cardelli, J. A., Clayton,
G. C., \& Mathis, J. S. 1989, \apj, 345, 245
\bibitem[Carone et al.(1996)]{car96} Carone, T. E. \etal\ 1996, \apj, 471, 737
\bibitem[Carroll, Press, \& Turner(1992)]{car92} Carroll, S. M., Press, W. H.,
\& Turner, E. L. 1992, \araa, 30, 499
\bibitem[Clements et al.(1996a)]{cle96a} Clements, D. L., Sutherland, W. J.,
Saunders, W., Efstathiou, G. P., McMahon, R. G., Maddox, S., Lawrence, A.,
\& Rowan-Robinson, M. 1996a, \mnras, 279, 459
\bibitem[Clements et al.(1996b)]{cle96b} Clements, D. L., Sutherland, W. J.,
McMahon, R. G., \& Saunders, W. 1996b, \mnras, 279,~477
\bibitem[Clements(2000)]{cle00} Clements, D. L. 2000, \mnras, 311, 833
\bibitem[Cohen et al.(1995)]{coh95} Cohen, M. H. \etal\ 1995, \apjl, 448, L77
\bibitem[Corbin(1997)]{cor97} Corbin, M. R. 1997, \apjs, 113, 245
\bibitem[Corbin \& Boroson(1996)]{cor96} Corbin, M. R., \& Boroson, T. A.
1996, \apjs, 107, 69
\bibitem[de Grijp, Miley, \& Lub(1987)]{deG87} de Grijp, M. H. K., Miley, G.
K., \& Lub, J. 1987, \aaps, 1987, 70, 95
\bibitem[Disney et al.(1995)]{dis95} Disney, M. J., \etal\ 1995, Nature, 376,
150
\bibitem[Egami et al.(1996)]{ega96} Egami, E., Iwamuro, F., Maihara, T.,
Oya, S., Cowie, L. L. 1996, \aj, 112, 73
\bibitem[Gehren et al.(1984)]{geh84} Gehren, T., Fried, J., Wehinger, P. A., \&
Wyckoff, S. 1984, \apj, 278, 11
\bibitem[Genzel et al.(1998)]{gen98} Genzel, R. \etal\ 1998, \apj, 498, 579
\bibitem[Glenn, Schmidt, \& Foltz(1994)]{gle94} Glenn, J., Schmidt, G. D., \&
Foltz, C. B. 1994, \apjl, 434, L47
\bibitem[Goodrich \& Miller(1994)]{goo94} Goodrich, R. W. \& Miller, J. S.
1994, \apj, 434, 82
\bibitem[Goodrich et al.(1996)]{goo96} Goodrich, R. W., Miller, J. S., Martel,
A., Cohen, M. H., Tran, H. D., Ogle, P. M., \& Vermeulen, R. C. 1996, \apjl,
456, L9
\bibitem[Grazian et al.(2000)]{gra00} Grazian, A., Cristiani, S., D'Odorico, V.,
Omizzolo, A., \& Pizzella, A. 2000, AJ, 119, 2540
\bibitem[Gregg et al.(2000)]{gre00} Gregg, M. D., Becker, R. H., Brotherton,
M. S., Laurent-Meuleisen, S. A., Lacy, M., White, R. L. 2000, \apj, 544, 142
\bibitem[Gunn(1979)]{gun79} Gunn, J. E. 1979 in {\em Active Galactic Nuclei},
ed.\ C. Hazard and S. Mitton (New York: Cambridge University Press), p. 213
\bibitem[Haas et al.(1998)]{haa98} Haas, M., Chini, R., Meisenheimer, K.,
Stickel, M., Lemke, D., Klaas, U., \& Kreysa, E. 1998, \apjl, 503, L109
\bibitem[Hazard et al.(1984)]{haz84} Hazard, C., Morton, D. C., Terlevich, R., \& McMahon, R. 1984, \apj, 282, 33
\bibitem[Heckman et al.(1984)]{hec84} Heckman, T. M., Bothun, G. D., Balick, 
B., Smith, E. P. 1984, \aj, 89, 958
\bibitem[Hernquist \& Quinn(1988)]{her88} Hernquist, L., \& Quinn, P. J. 1988, \apj, 331, 682
\bibitem[Hernquist \& Quinn(1989)]{her89} Hernquist, L., \& Quinn, P. J. 1989, \apj, 342, 1
\bibitem[Hernquist \& Spergel(1992)]{her92} Hernquist, L., \& Spergel, D. N.
1992, \apj, 399, 117
\bibitem[Hes, Barthel, \& Fosbury(1993)]{hes93} Hes, R., Barthel, P. D., \&
Fosbury, R. A. E. 1993, Nature, 362, 326
\bibitem[Hickson \& Hutchings(1987)]{hic87} Hickson, P. \& Hutchings, J. B.
1987, \apj, 312, 518
\bibitem[Hines et al.(1999b)]{hin99b} Hines, D. C., Low, F. J., Thompson, R.
J., Weymann, R. J., \& Storrie-Lombardi, L. 1999, \apj, 512, 140
\bibitem[Hines \& Schmidt(1997)]{hin97} Hines, D. C., \& Schmidt, G. D. 1997,
in Mass Ejection from Active Galactic Nuclei, ASP Conference Series 128,
ed.\ N. Arav, I. Shlosman, \& R. J. Weymann, p.~99
\bibitem[Hines et al.(1995)]{hin95a} Hines, D. C., Schmidt, G. D., Smith, P.
S., Cutri, R. M., Low, F. J. 1995, \apjl, 450, L1
\bibitem[Hines et al.(1999a)]{hin99a} Hines, D. C., Schmidt, G. D., Wills,
B. J., Smith, P. S., \& Sowinski, L. G. 1999, \apj, 512, 145
\bibitem[Hines \& Wills(1995)]{hin95b} Hines, D. C., \& Wills, B. J. 1995,
\apjl, 448, L69
\bibitem[Hodapp et al.(1996)]{hod96} Hodapp, K.-W., et al.\ 1996, New Astronomy,
1, 177
\bibitem[Houck et al.(1984)]{hou84} Houck, J. R. \etal\ 1984, \apjl, 278, L63
\bibitem[Hughes et al.(2000)]{hug00} Hughes, D. H., Kukula, M. J., Dunlop,
J. S., Boroson, T. 2000, \mnras, 316, 204
\bibitem[Hutchings et al.(1982a)]{hut82a} Hutchings, J. B., Campbell, B.,
Gower, A. C., Crampton, D., \& Morris, S. C. 1982a, \apj, 262, 48
\bibitem[Hutchings \& Crampton(1990)]{hut90} Hutchings, J. B., \& Crampton, 
D. 1990, \aj, 99, 37
\bibitem[Hutchings et al.(1984)]{hut84} Hutchings, J. B., Crampton, D.,
Campbell, B., Duncan, D., \& Glendenning, B. 1984, \apjs, 55, 319
\bibitem[Hutchings et al.(1981)]{hut81} Hutchings, J. B., Crampton, D.,
Campbell, B., Pritchet, C. 1981, \apj, 247, 743
\bibitem[Hutchings et al.(1994)]{hut94} Hutchings, J. B., Holtzman, J.,
Sparks, W. B., Morris, S. C., Hanisch, R. J., \& Mo, J. 1994, \apjl, 429, L1
\bibitem[Hutchings, Johnson, \& Pyke(1988)]{hut88a} Hutchings, J. B.,
Johnson, I., \& Pyke, R. 1988, \apjs, 66, 361
\bibitem[Hutchings \& Neff(1988)]{hut88b} Hutchings, J. B. \& Neff, S. G. 
1988, \aj, 96, 1575
\bibitem[Hutchings \& Neff(1992)]{hut92} Hutchings, J. B., \& Neff, S. G.
1992, \aj, 104, 1
\bibitem[Hutsem\'{e}kers, Lamy, \& Remy(1998)]{hut98} Hutsem\'{e}kers, D.,
Lamy, H., \& Remy, M. 1998, \aap, 340, 371
\bibitem[Joly(1991)]{jol91} Joly, M. 1991, \aap, 242, 49
\bibitem[Joseph(1999)]{jos99} Joseph, R. D. 1999, \apss, the
  Ringberg Workshop, ``Ultraluminous Galaxies: Monsters or Babies'' (Ringberg
  Castle, September 1998), 266, p. 321
\bibitem[Junkkarinen, Hewitt, \& Burbidge(1991)]{jun91} Junkkarinen, V., 
Hewitt, A., \& Burbidge, G. 1991, \apjs, 77, 203
\bibitem[Kim \& Sanders(1998)]{kim98} Kim, K.-C. \& Sanders, D. B. 1998,
\apjs, 119, 41
\bibitem[Klaas et al.(1997)]{kla97} Klaas, U., Haas, M., Heinrichsen, I., \&
Schutz, B. 1997, \aap, 325, L21
\bibitem[Kleinmann et al.(1988)]{kle88} Kleinmann, S. G., Hamilton, D.,
Keel, W. C., Wynn-Williams, C. G., Eales, S. A., Becklin, E. E., \&
Kuntz, K. D. 1988, \apj, 328, 161
\bibitem[Lanzetta, Turnshek, \& Sandoval(1993)]{lan93} Lanzetta, K. M.,
Turnshek, D. A., \& Sandoval, J. 1993, \apjs, 84, 109
\bibitem[Lim \& Ho(1999)]{lim99} Lim, J. \& Ho, P. 1999, \apjl, 510, L7
\bibitem[L\'{\i}pari(1994)]{lip94} L\'{\i}pari, S. 1994 \apj, 436, 102
\bibitem[Lonsdale, Smith, \& Lonsdale(1995)]{lon95} Lonsdale, C. J., Smith,
H. E., \& Lonsdale, C. J. 1995, \apj, 438, 632
\bibitem[Lonsdale, Smith, \& Lonsdale(1993)]{lon93} Lonsdale, C. J., Smith,
H. E., \& Lonsdale, C. J. 1993, \apjl, 405, L9
\bibitem[Low et al.(1989)]{low89} Low, F. J., Cutri, R. M., Kleinmann, S. G.,
\& Huchra, J. P.  1989, \apjl, 340, L1
\bibitem[Low et al.(1988)]{low88} Low, F. J., Huchra, J. P., Kleinmann, S. G.,
Cutri, R. M. 1988, \apjl, 327, L41
\bibitem[Lutz et al.(1998)]{lut98} Lutz, D., Spoon, H. W. W., Rigopoulou, D.,
Moorwood, A. F. M., Genzel, R. 1998, \apjl, 505, L103
\bibitem[Lutz, Veilleux, \& Genzel(1999)]{lut99} Lutz, D., Veilleux, S., \& Genzel,
R. 1999, \apjl, 517, L13
\bibitem[Lynden-Bell(1969)]{lyn69} Lynden-Bell, D. 1969, Nature, 223, 690
\bibitem[Lynds \& Toomre(1976)]{lyn76} Lynds, R., \& Toomre, A. 1976, \apj,
209, 382
\bibitem[Mackenty \& Stockton(1984)]{mac84} MacKenty, J. W., \& Stockton, A.
N. 1984, \apj, 283, 64
\bibitem[Malin(1979)]{mal79} Malin, D. F. 1979, Nature, 277, 279
\bibitem[Malin \& Carter(1980)]{mal80} Malin, D. F., \& Carter, D. 1980,
Nature, 285, 643
\bibitem[McIntosh et al.(1999)]{mci99} McIntosh, D. H., Rieke, M. J., Rix,
H.-W., Foltz, C. B., \& Weymann, R. J. 1999, \apj, 514, 40
\bibitem[McLeod \& Rieke(1994)]{mcl94} McLeod, K. K., \& Rieke, G. H. 1994,
\apj, 431, 137
\bibitem[McLure et al.(1999)]{mcl99} McLure, R. J., Kukula, M. J., Dunlop,
J. S., Baum, S. A., O'Dea, C. P., \& Hughes, D. H. 1999, \mnras, 308, 377
\bibitem[Meylan(1993)]{mey93} Meylan, G. 1993, in {\em The Globular Cluster
Galaxy Connection,} ed. G. H. Smith \& J. P. Brodie, ASP Conference Series,
vol.48, p.~588
\bibitem[Mihos \& Hernquist(1994)]{mih94} Mihos, J. C. \& Hernquist, L. 1994,
425, L13
\bibitem[Mihos \& Hernquist(1996)]{mih96} Mihos, J. C. \& Hernquist, L. 1996,
\apj, 464, 641
\bibitem[Nakagawa et al.(1999)]{nak99} Nakagawa, T., Kii, T., Fujimoto, R.,
Miyazaki, T., Inoue, H., Ogasaka, Y., Kawabe, R. 1999, Ap\&SS, 266, 43
\bibitem[Neugebauer, Soifer, \& Miley(1985)]{neu85} Neugebauer, G., Soifer,
B. T., \& Miley, G. K. 1985, \apjl, 295, L27
\bibitem[Neugebauer et al.(1986)]{neu86} Neugebauer, G., Miley, G. K., Soifer,
B. T., \& Clegg, P. E. 1986, \apj, 308, 815
\bibitem[Nolan et al.(2000)]{nol00} Nolan, L. A.,  Dunlop, J. S., Kukula,
M. J., Hughes, D. H., Boroson, T., \& Jimenez, R. 2000, \mnras, in press
[astro-ph/0004325]
\bibitem[Oke et al.(1995)]{oke95} Oke, J. B., et al. 1995, \pasp, 107, 375
\bibitem[Sanders(1999)]{san99} Sanders, D. B. 1999, \apss,
   the Ringberg Workshop, ``Ultraluminous Galaxies: Monsters or Babies''
   (Ringberg Castle, September 1998), 266, p. 331
\bibitem[Sanders \& Mirabel(1996)]{san96} Sanders, D. B. \& Mirabel, I. F. 1996, \araa, 34, 749
\bibitem[Sanders et al.(1989)]{san89} Sanders, D. B., Phinney, E. S.,
Neugebauer, G., Soifer, B. T., \& Matthews, K. 1989, \apj, 347, 29
\bibitem[Sanders et al.(1986)]{san86} Sanders, D. B., Scoville, N. Z.,
Young, J. S., Soifer, B. T., Schloerb, F. P., Rice, W. L., \& Danielson, G. E.
1986, \apjl, 305, 45
\bibitem[Sanders et al.(1988)]{san88} Sanders, D. B., Soifer, B. T., Elias,
J. H., Madore, B. F., Matthews, K., Neugebauer, G., \& Scoville, N. Z. 1988,
\apj, 325, 74
\bibitem[Sansom, Hibbard, \& Schweizer(2000)]{san00} Sansom, A. E., Hibbard,
J. E., Schweizer, F. 2000, \aj, 120, 1946
\bibitem[Sargent(1968)]{sar68} Sargent, W. L. W. 1968, \apjl, 151, L31
\bibitem[Scalo(1986)]{sca86} Scalo, J. M. 1986, Fund. Cosmic Phys., 11, 1
\bibitem[Schinnerer, Eckart, \& Tacconi(1998)]{scn98} Schinnerer, E., Eckart,
A., \& Tacconi, L. J. 1998, \apj, 500, 147
\bibitem[Schlegel, Finkbeiner, \& Davis(1998)]{sch98} Schlegel, D. J.,
Finkbeiner D. P., \& Davis, M. 1998, \apj, 500, 525
\bibitem[Schmidt \& Green(1983)]{sch83} Schmidt, M., \& Green, R. F. 1983,
\apj, 269, 352
\bibitem[Schweizer \& Seitzer(1988)]{sch88} Schweizer, F., \& Seitzer, P. 1988,
\apj, 328, 88
\bibitem[Schweizer \& Seitzer(1992)]{sch92} Schweizer, F., \& Seitzer, P. 1992,
\aj, 104, 1039
\bibitem[Schweizer et al.(1990)]{sch90} Schweizer, F., Seitzer, P., Faber,
S. M., Burstein, D., Dalle Ore, C. M., \& Gonzalez, J. J. 1990, \apj, 364, L33
\bibitem[Shields(1996)]{shi96} Shields, G. A. 1996, \apjl, 461, L9
\bibitem[Simpson(1998)]{sim98} Simpson, C. 1998, \mnras, 297, L39
\bibitem[Smith, Lonsdale, \& Lonsdale(1998)]{smi98} Smith, H., E., Lonsdale,
C. J., Lonsdale, C. J. 1998, \apj, 492, 137
\bibitem[Soifer et al.(1984a)]{soi84a} Soifer, B. T. \etal\ 1984a, \apjl, 278,
L71
\bibitem[Soifer et al.(1984b)]{soi84b} Soifer, B. T., Neugebauer, G., Helou,
G., Lonsdale, C. J., Hacking, P., Rice, W., Houck, J. R., Low, F. J., \&
Rowan-Robinson, M.  1984b, \apjl, 283, L1
\bibitem[Soifer et al.(1986)]{soi86} Soifer, B. T., Sanders, D. B.,
Neugebauer, G., Danielson, G. E., Lonsdale, C. J., Madore, B. F., Persson,
S. E. 1986, \apjl, 303, L41
\bibitem[Soifer et al.(1987)]{soi87} Soifer, B. T., Sanders, D. B., Madore,
B. F., Neugebauer, G., Danielson, G. E., Elias, J. H., Lonsdale, C. J.,
Rice, W. L. 1987, \apj, 320, 238.
\bibitem[Sowinski, Schmidt, \& Hines(1997)]{sow97} Sowinski, L. G., Schmidt,
G. D., \& Hines, D. C. 1997, in Mass Ejection from Active Galactic Nuclei,
ASP Conference Series 128, ed.\ N. Arav, I. Shlosman, \& R. J. Weymann, p.~305
\bibitem[Sprayberry \& Foltz(1992)]{spr92} Sprayberry, D. \& Foltz, C. B.
1992, \apj, 390, 39
\bibitem[Stockton(1982)]{sto82} Stockton, A. 1982, \apj, 257, 33
\bibitem[Stockton(1990)]{sto90} Stockton, A. 1990, in {\em Dynamics and
Interactions of Galaxies}, ed.\ R. Wielen, (Springer-Verlag: Berlin), 440
\bibitem[Stockton, Canalizo, \& Close(1998)]{sto98} Stockton, A., Canalizo, G.,
\& Close, L. 1998, \apjl, 500, L121 [SCC98]
\bibitem[Stockton \& MacKenty(1987)]{sto87} Stockton, A., \& MacKenty 1987,
\apj, 316, 584
\bibitem[Stockton \& Ridgway(1991)]{sto91} Stockton, A., \& Ridgway, S. E.
1991, \aj, 102, 488
\bibitem[Surace(1998)]{sur98a} Surace, J. A., 1998, Ph.D. thesis, University of
Hawaii
\bibitem[Surace \& Sanders(2000)]{sur00} Surace, J. \& Sanders, D. B. 2000,
\aj, 120, 604
\bibitem[Surace et al.(1998)]{sur98b} Surace, J. A., Sanders, D. B., Vacca,
W. D., Veilleux, S., \& Mazzarella, J. M. 1998, \apj, 492, 116
\bibitem[Taniguchi et al.(1988)]{tan88} Taniguchi, T., Kawara, K., Nishida, M.,
Tamura, S., Nishida, M. T. 1988, \apj, 95, 1378
\bibitem[Terlevich et al.(1992)]{ter92} Terlevich, R., Tenorio-Tagle, G.,
Franco, J., \& Melnick, J 1992, \mnras, 255, 713
\bibitem[Toomre(1977)]{too77} Toomre, A. 1977, in The Evolution of Galaxies
and of Stellar Populations, eds.\ B. M. Tinsley \& R. B. Larson (Yale
Observatory, New Haven), p. 401
\bibitem[Toomre \& Toomre(1972)]{too72} Toomre, A., \& Toomre, J. 1972, \apj,
178, 623
\bibitem[Tran et al.(1999)]{tra99} Tran, H. D., Brotherton, M. S., Stanford,
S. A., van Breugel, W., Dey, A., Stern, D., \& Antonucci, R. 1999, \apj, 516, 85
\bibitem[Tran, Cohen, \& Villar-Martin(2000)]{tra00}Tran, H. D., Cohen, M. H.,
\& Villar-Martin, M. 2000, \aj, in press [astro-ph/0004383]
\bibitem[Turnshek et al.(1997)]{tur97} Turnshek, D. A., Monier, E. M., Sirola,
C. J., \& Espey, B. R.  1997, \apj, 476, 40
\bibitem[Vader et al.(1987)]{vad87} Vader, J. P., Da Costa, G. S., Frogel,
J. A., Heisler, C. A., Simon, M. 1987, \aj, 94, 847
\bibitem[Vader \& Simon(1987)]{vs87} Vader, J. P., \& Simon, M. 1987,
Nature, 327, 304
\bibitem[Veilleux(1999)]{vei99a} Veilleux, S. 1999, in Galaxy Interactions at Low
and High Redshift (IAU Symp.~186), ed.~J. E. Barnes \& D. B. Sanders
(Dordrecht: Kluwer), p.~295
\bibitem[Veilleux \& Osterbrock(1987)]{vei87} Veilleux, S. \& Osterbrock,
D. E. 1987, \apjs, 63, 295
\bibitem[Veilleux, Sanders, \& Kim(1997)]{vei97} Veilleux, S., Sanders, D. B.,
\& Kim, D.-C. 1997, \apj, 484, 92
\bibitem[Veilleux, Kim, \& Sanders(1999)]{vei99} Veilleux, S., Kim, D.-C.,
\& Sanders, D. B. 1999, \apj, 522, 113
\bibitem[V\'{e}ron-Cetty \& V\'{e}ron(1996)]{ver96} V\'{e}ron-Cetty, M.P. \&
V\'{e}ron, P. 1996, ESO Sci. Rep. 17, 1
\bibitem[Voit, Weymann, \& Korista(1993)]{voi93} Voit, G. M., Weymann, R. J., \&
Korista, K. T. 1993, \apj, 413, 95
\bibitem[Weymann(1997)]{wey97} Weymann, R. 1997, in Mass Ejection from Active
Galactic Nuclei, ASP Conference Series 128, ed.\ N. Arav, I. Shlosman, \& R. J.
Weymann, p.\ 3
\bibitem[Weymann et al.(1991)]{wey91} Weymann, R. J., Morris, S.L., Foltz,
C.B., \& Hewett, P.C. 1991, \apj, 373, 23
\bibitem[Wills(1996)]{wil96a} Wills, B. J. 1996, in Jets from Stars to Active
Galactic Nuclei, ed.\ W. Kundt (Berlin: Springer), 213
\bibitem[Wills \& Brotherton(1996)]{wil96b} Wills, B. J., \& Brotherton, M. S.
1996, in Jets from Stars to Active Galactic Nuclei, ed. W. Kundt, (Berlin:
Springer), p.~203
\bibitem[Wills \& Hines(1997)]{wil97} Wills, B. J., \& Hines, D. C. 1997, in
Mass Ejection from Active Galactic Nuclei, ASP Conference Series 128, ed.\ N.
Arav, I. Shlosman, \& R. J. Weymann, p.~99
\bibitem[Wills et al.(1992)]{wil92} Wills, B. J., Wills, D., Evans, N. J., II,
Natta, A, Thompson, K. L., Breger, M., \& Sitko, M. L. 1992, \apj, 400, 96
\bibitem[Wyckoff, Gehren, \& Wehinger(1981)]{wyc81} Wyckoff, S., Gehren, T.,
\& Wehinger, P. A. 1981, \apj, 247, 750
\bibitem[Zheng et al.(1999)]{zhe99} Zheng, Z., \etal\ 1999, \aap, 349, 735
\end{thebibliography}
\end{document}